\documentclass[aps,prd,showpacs,showkeys,twocolumn,groupedaddress]{revtex4}

\usepackage{amsmath}
\usepackage{amsfonts}
\usepackage{amssymb}
\usepackage{latexsym}
\usepackage{graphicx}
\usepackage{float}
\usepackage{wrapfig}
\usepackage{epsfig}
\usepackage{diagbox}
\usepackage{placeins}
\usepackage{hyperref} 

\newcommand{\be}{\begin{eqnarray}}
\newcommand{\ee}{\end{eqnarray}}

\newcommand{\bmat}{\left ( \begin{array}{cc} 
}
\newcommand{\emat}{\end{array} \right )}

\newcommand{\Ort}{{\rm O\,}}

\newcommand{\USp}{{\rm USp\,}}
\newcommand{\U}{{\rm U\,}}
\newcommand{\SU}{{\rm SU\,}}

\newcommand{\tr}{{\rm tr\,}}
\newcommand{\sign}{{\rm sign\,}}
\newcommand{\eins}{\leavevmode\hbox{\small1\kern-3.8pt\normalsize1}}

\begin{document}
 \title{Shift of Symmetries of Naive Fermions in QCD-like Lattice Theories}
\author{Mario Kieburg}
\email{mkieburg@physik.uni-bielefeld.de}
\author{Tim R. W\"urfel}
\email{twuerfel@physik.uni-bielefeld.de}
\date{\today}
\affiliation{Faculty of Physics, Bielefeld University, PO-Box 100131, 33501 Bielefeld, Germany}

\begin{abstract}
 We study the global symmetries of naive lattices Dirac operators in QCD-like theories in any dimension larger than two. In particular we investigate how the chosen number of lattice sites in each direction affects the global symmetries of the Dirac operator. These symmetries are important since they do not only determine the infra-red spectrum of the Dirac operator but also the symmetry breaking pattern and, thus, the lightest pseudo-scalar mesons. We perform the symmetry analysis and discuss the possible zero modes and the degree of degeneracy of the lattice Dirac operators. Moreover we explicitly identify a ``reduced" lattice Dirac operator which is the naive Dirac operator apart from the degeneracy. We verify our predictions by comparing Monte Carlo simulations of QCD-like theories in the strong coupling limit with the corresponding random matrix theories.
\end{abstract}

\maketitle

\section{Introduction}\label{sec:intro}

QCD-like theories describing the strong interaction between quarks and gluons are highly involved due to their non-linear field equations. This statement also applies to QCD-like theories beyond the standard model like theories with technicolor~\cite{technicolor} or supersymmetry~\cite{SUSYQCD}. Therefore quite often only numerical simulations remain as a tool to study the theory as a whole. Here, one encounters two crucial modifications of the continuum theory to overcome certain problems.

First, the time axis is Wick rotated to circumvent the sign problem resulting from a real time. Thus, the Yang-Mills action is replaced by a Euclidean Wilson gauge action or a Symanzik improved version of it. Furthermore, the Dirac operator becomes a hypercubic lattice Dirac operator which is a finite dimensional matrix on a lattice of finite volume $V$. There are several lattice discretizations of the QCD-Dirac operator. To name only the most prominent, the Dirac operators of staggered fermions~\cite{staggered}, of Wilson fermions~\cite{Wilson}, of twisted mass fermions~\cite{Twisted}, of overlap fermions~\cite{Overlap} and of domain wall fermions~\cite{Domain}.

The simplest version of a lattice Dirac operator is the naive Dirac operator on a cubic lattice with periodic and anti-periodic boundary conditions. The Dirac operator of staggered fermions (without rooting) is a particular version of the naive Dirac operator. It is the non-degenerate part of the naive Dirac operator on a cubic lattice were each direction consists of an even number of lattice sites.

The second problem to be solved in lattice QCD concerns the continuum limit. It is well-known that the global symmetries of staggered fermions in four dimensions do not necessarily agree with those of the continuum theory for QCD. For example the theories  with two colors and the fermions in the fundamental representations or with arbitrary colors in the adjoint representation have this problem, see~\cite{4DstagQCD.a,4DstagQCD.b}. This problem was also found in three~\cite{3DstagQCD.a,3DstagQCD.b} and in two dimensions~\cite{2DstagQCD}. Especially in two dimensions the reason for the change of the global symmetries was recently analyzed in detail in~\cite{2DstagQCD}. In the present work we aim at a generalization of this discussion to arbitrary space-time dimensions.

The global symmetries are manifest in the lowest eigenvalues of the Dirac operator. In the phase of spontaneous breaking of chiral symmetry the spectral gap is closed and chiral perturbation theory applies. The order parameter is the chiral condensate $\Sigma$ which is given by the Banks-Casher formula~\cite{BanksCasher} in terms of the level density $\rho$ of the Dirac operator,
\begin{equation}\label{BCrelation}
\Sigma=|\langle\bar{\psi}\psi\rangle|=\lim_{a\to0}\lim_{m\to0}\lim_{V\to\infty}\frac{\pi}{V}\int\frac{2m\rho(\lambda)d\lambda}{m^2+\lambda^2}
\end{equation}
The order of the limits is crucial. The Banks-Casher formula ~\eqref{BCrelation} is only true for Dirac operators exhibiting a chiral symmetry. In three dimensions one has to consider the following condensate
\begin{equation}\label{BCrelation-nonchiral}
\Sigma_{\rm non\chi}=|\langle\bar{\psi}\tau_3\psi\rangle|=\lim_{a\to0}\lim_{m\to0}\lim_{V\to\infty}\frac{\pi}{V}\int\frac{2m\rho(\lambda)d\lambda}{m^2+\lambda^2}.
\end{equation}
Note that the expression in terms of the level density is the same in both equations because we assumed for the latter an even number of dynamical quarks. Then, half the number of the quarks contribute a chiral condensates weighted with a plus sign and the other half with a minus sign.

From the expressions~\eqref{BCrelation} and \eqref{BCrelation-nonchiral} we recognize that only those eigenvalues of the Dirac operator of order $\mathcal{O}(1/V)$ are important for the spontaneous symmetry breaking. The corresponding modes are intimately related with those pseudo-scalar mesons whose Compton wavelength is larger than the size of the lattice. Then the kinetic modes factorize from the zero-momentum modes~\cite{GasserLeutwyler,LeutwylerSmilga} and can be integrated out. The remaining effective Lagrangian for the zero-momentum modes is shared with Gaussian random matrix models~\cite{ShuVer93,Verthreefold}. This theoretical prediction was several times numerically verified, for example  for the microscopic level density
\begin{equation}\label{microleveldens}
\rho_{\rm micro}(x)=\lim_{V\to\infty}\frac{1}{\Sigma V}\rho\left(\frac{x}{\Sigma V}\right),
\end{equation}
see \cite{4DstagQCD.a}.

For a long time it is known that QCD-like theories may yield different patterns of spontaneous symmetry breaking and, hence, different kinds of Goldstone bosons, e.g. see~\cite{Peskin,Preskill,Verthreefold}. In \cite{contclass} the symmetry breaking patterns were derived for an arbitrary dimension larger than two. For QCD-like theories in one and two dimensions the Mermin-Wagner-Coleman theorem~\cite{MerminWagner,coleman} forbids a spontaneous breaking of global symmetries, though two-dimensional theories seem to be at the borderline, see~\cite{2DstagQCD}. The authors of~\cite{contclass} found that the Bott-periodicity~\cite{Bott} satisfied by the $\gamma$-matrices carries over to the symmetry breaking pattern and, thus, to the effective Lagrangian for the pseudo-scalar mesons. A similar classification was also found for topological insulators~\cite{topInsu}.  All in all there are ten symmetry breaking patterns associated to QCD-like Dirac operators. They correspond to the Cartan classification of Hermitian random matrices~\cite{threefoldway,tenfoldway} and each of these ten random matrix ensembles yields the effective Lagrangian of the pseudo-scalar mesons in lowest order. 

Marrying the two observations that the partition of the lattice and the space-time dimension affect the 
global symmetries and, hence, the symmetry breaking pattern, one may ask the question what their combined impact is for an arbitrary QCD-like theory. Answering this question is the main task of the present work. For this purpose we pursue similar ideas as in~\cite{2DstagQCD}.

We first review the symmetry analysis of the continuum theory~\cite{contclass}, in section~\ref{sec:continuum}. In this section  we also recall the properties of Clifford algebras built up by the $\gamma$-matrices and the different kinds of anti-unitary operators which are at the heart of the  classification of global symmetries. In section~\ref{sec:lattice} we very briefly review what the naive discretization explicitly looks like and what the corresponding symmetry operations on a lattice are. Especially we  discuss the artificial symmetry operations which arise when one or more directions have an even partition of lattice sites. Those additional symmetry operators anti-commute  with the lattice Dirac operator and build a Clifford algebra themselves. They are the origin for the change of symmetries which is analyzed in section~\ref{sec:symmetry}. In section~\ref{sec:sympat} we discuss the symmetry breaking patterns and the number of zero modes and in section~\ref{sec:MC-sim} we derive an explicit representation of the non-degenerate part of the lattice Dirac operator. This ``reduced" Dirac operator is the staggered Dirac operator when each direction of the lattice contains an even number of sites. We also perform lattice simulations in the strong coupling limit and compare it with random matrix ensembles predicted by our analysis. Those comparisons are shown in figures~\ref{fig1} and \ref{fig2}. The random matrix results employed in these comparisons are recalled in appendix~\ref{sec:RMT}. In section~\ref{sec:conclusio} we summarize our results and discuss some important implications. The tables~\ref{tab:cont-class}, \ref{tab:lat-class} and \ref{tab:RMT} shall help to get an overview of the change of symmetries when varying the space-time dimension, the number of directions with an even partition and the representation of the gauge group. We want to underline that we deal all QCD-like gauge theories on equal footing and do not restrict ourselves to the gauge group ${\rm SU}(N_{\rm c})$.

In contrast to standard literature we do not apply Einstein's summation convention throughout the present work because of computational reasons.

\section{Dirac Operator of QCD-like Theories in Euclidean Continuum}\label{sec:continuum}

We consider the Euclidean massless Dirac Operator of a QCD-like gauge theory in $d$-dimensions, i.e.
\begin{equation}\label{def:Dirac-cont}
\mathcal{D}=\sum_{\mu=1}^d\left(\partial_\mu\eins_{d_r}+\imath A_\mu(x)\right)\gamma_\mu
\end{equation}
with $\imath$  the imaginary unit. The vector fields $\imath A_\mu(x)\in r(\mathfrak{g})$, with $x$ a point in space-time, are elements in an irreducible representation $r$ of a finite-dimensional compact Lie-algebra $\mathfrak{g}$, in particular $A_\mu^\dagger=A_\mu$ is anti-Hermitian. The dimension of this representation is denoted by $d_r$ and should not be confused with the space-time dimension $d$. In QCD the Lie-algebra is $\mathfrak{g}={\rm su}(3)$. The dimension $d_r$ is emphasised by the identity matrix $\eins_{d_r}$ multiplied with the partial derivative $\partial_\mu$ in the $\mu$'th direction. Since we consider Euclidean space-time the generalized $\gamma$-matrices are given by the Clifford algebra~\cite{Clifford}
\begin{equation}\label{Clifford}
 [\gamma_\mu,\gamma_\nu]_+=2\delta_{\mu\nu}\eins_{2^{\lfloor d/2\rfloor}},\ \tr\gamma_\mu=0\text{ and }\gamma_\mu^\dagger=\gamma_\mu
\end{equation}
with $[.,.]_+$ the anti-commutator and $\gamma_\mu$ being Hermitian $2^{\lfloor d/2\rfloor}\times 2^{\lfloor d/2\rfloor}$ matrices. The function $\lfloor d/2\rfloor$ is the floor function yielding the largest integer equal or smaller than $d/2$.

The Clifford algebra above generates via multiplication of algebra elements and scalars as well as addition of algebra elements the fundamental representation of the Lie-algebra of the unitary group $\U(2^{\lfloor d/2\rfloor})$. This fact will become helpful later on.

We now recall some well-known facts about the Clifford algebra with $d$ generators, see also \cite{Clifford}. The generalized $\gamma$-matrices can be represented in terms of the three Pauli-matrices $\sigma_i$, $i=1,2,3$. However we will state our results basis independently.  We recall that for odd dimension $d$ (odd number of generators of the Clifford algebra) there is no chiral symmetry while for even dimensions the matrix
\begin{equation}\label{def:gamma5}
\gamma^{(5)}=\imath^{-d(d-1)/2}\prod_{j=1}^d\gamma_j=\imath^{-d(d-1)/2}\gamma_1\gamma_2\cdots\gamma_d
\end{equation}
anti-commutes with all generators such that there is a chiral symmetry. Moreover, the generators of the Clifford algebra always satisfy an anti-unitary symmetry
\begin{equation}\label{anti-unitary}
[C,\imath^{d(d-1)/2}\gamma_\nu]_-=0\text{ and }[C,\imath^{d(d-1)/2}\gamma^{(5)}]_-=0
\end{equation}
with $[.,.]_-$ the commutator. The operator $C$ can be explicitly written as
\begin{equation}\label{def:C}
 C=K\chi=\left\{\begin{array}{cl} K\imath^{m}\prod_{j=1}^{2m}\gamma_{2j}, & d=4m,4m+1,\\ K\imath^{m+1}\prod_{j=1}^{2m+2}\gamma_{2j-1}, & d=4m+2,4m+3, \end{array}\right.
\end{equation}
with $K$ the complex conjugation operator and $\gamma_{d+1}=\gamma^{(5)}$ for $d\in2\mathbb{N}+1$. It satisfies
\begin{equation}\label{C-square}
 C^2=(-1)^{(d+2)(d+1)d(d-1)/8}\eins_{2^{\lfloor d/2\rfloor}}
\end{equation}
which is the origin for the Bott periodicity~\cite{Bott} of Clifford algebras. The matrix $\chi$, see eq.~\eqref{def:C}, is a unitary, Hermitian matrix which is either symmetric or anti-symmetric depending on the sign of $C^2$.

The anti-unitary symmetry~\eqref{anti-unitary} has to be combined with the one which might be satisfied by the vector fields $A_\mu$. Note that the partial derivatives are anti-Hermitian and real, i.e. $\partial_\mu=-\partial_\mu^\dagger=K\partial_\mu K$.  Also the vector fields $\imath A_\mu$ are anti-Hermitian, i.e. $(\imath A_\mu)^\dagger=-\imath A_\mu$. Additionally they might satisfy an anti-unitary symmetry,
\begin{equation}\label{anti-unitary-A}
[K\zeta,\imath A_\mu]_+=0\text{ and }(K\zeta)^2=\zeta^*\zeta=\pm\eins
\end{equation}
with $\zeta$ a unitary matrix. If there is no such symmetry then the representation $r$ is called complex. Examples are the fundamental representations of the gauge groups ${\rm SU}(N_{\rm c}\geq3)$ and  ${\rm U}(N_{\rm c}\geq2)$.  When there is an anti-unitary operator $K\zeta$, the representation $r$ is called real or quaternion when $(K\zeta)^2=\eins$ or $(K\zeta)^2=-\eins$, respectively. Examples for real representations are the fundamental representation of ${\rm SO}(2N_{\rm c}+1\geq 5)$ and ${\rm SO}(2N_{\rm c}\geq 8)$ and  the adjoint representation of any compact Lie-algebra. The fundamental representation of ${\rm SU}(N_{\rm c}=2)={\rm USp}(2N_{\rm c}=2)$ and in general the unitary symplectic group ${\rm USp}(2N_{\rm c})$ are examples for quaternion representations.

The symmetry discussion above for the continuum Dirac operator is summarized in table~\ref{tab:cont-class} and was already performed in \cite{contclass}. For this purpose, we defined the charge conjugation operator
\begin{equation}
\mathcal{C}=K\zeta \chi
\end{equation}
which only exists for real and quaternion representations of the chosen gauge group. In  table~\ref{tab:cont-class} we also point out the symmetry breaking pattern, the symmetry class via the Cartan classification scheme~\cite{threefoldway,tenfoldway}, and the random matrix theory describing the infra-red energy spectrum of the Dirac operator. Note that all ten symmetry classes of Hermitian operators can be found as it is the case for topological insulators~\cite{topInsu}.

\section{Naive and Staggered Lattice Dirac Operator}\label{sec:lattice}

On the lattice the Hilbert space $\mathcal{H}=\mathbb{C}^{d_r}\otimes\widehat{V}\otimes\mathbb{C}^{\lfloor d/2\rfloor}$ is finite dimensional. It consists of three parts. The first part is the color space $\mathbb{C}^{d_r}$ where the gauge group acts on. The second one is the cubic lattice $\widehat{V}=\bigotimes_{j=1}^d\mathbb{C}^{L_j}$ with the volume $V=\prod_{j=1}^dL_j$ and $L_\mu\in\mathbb{N}$ being the number of lattice sites in the direction $\mu$. The third part is the spinor space $\mathbb{C}^{\lfloor d/2\rfloor}$. Thus the dimension of the Hilbert space is in total $d_{\mathcal{H}}=2^{\lfloor d/2\rfloor}d_r V$.

In the naive discretization the covariant derivative is replaced by the difference of the transfer matrix $T_\mu$ and its Hermitian adjoint, i.e.
\begin{equation}\label{cov-der}
 \partial_\mu\eins_{d_r}+\imath A_\mu(x)\longrightarrow T_\mu-T_\mu^{\dagger}.
\end{equation}
The transfer matrix $T_\mu$ acts as follows on a state $|\psi(x)\rangle\in \mathcal{H}$ at a fixed lattice site $x=(x_1,\ldots,x_d)\in\widehat{V}$,
\begin{equation}\label{def:transfer}
 T_\mu|\psi(x)\rangle=(-1)^{\delta_{jd}\delta_{x_dL_d}}U_\mu(x)|\psi(x+e_\mu)\rangle,
\end{equation}
where $e_\mu$ is the $d$-dimensional vector with a $1$ at the position $\mu$ and otherwise zero and $x_j\in\mathbb{Z}_{L_j}$. The matrices $U_\mu(x)$ are elements of the representation $r(\mathcal{G})$ with $\mathcal{G}$ the gauge group. This representation satisfies the same anti-unitary symmetry as the Lie algebra $\mathfrak{g}$ if existent. The sign in eq.~\eqref{def:transfer} reflects the boundary conditions which are periodic in the spatial directions $\mu=1,\ldots,d-1$ and anti-periodic for the temporal direction $\mu=d$. We want to emphasize that the results of the symmetry analysis will be independent of the periodic or anti-periodic boundary conditions. Hence the sign only plays a minor role.
The naive lattice Dirac operator is given by
\begin{equation}\label{def:lat-Dirac}
 D=\sum_{\mu=1}^d (T_\mu-T_\mu^{\dagger})\gamma_\mu.
\end{equation}

As already found for two dimensions, see \cite{2DstagQCD}, depending on the number of lattice sites in a fixed direction $\mu$ there might be an operator anti-commuting with the lattice Dirac operator~\eqref{def:lat-Dirac}. Suppose the number $L_\mu$ is even. Then we can consider the operator
\begin{equation}\label{def:gamma-lat}
 \Gamma_\mu|\psi(x)\rangle=(-1)^{x_\mu}|\psi(x)\rangle
\end{equation}
which assigns to each even lattice site (according to the parity in the direction $\mu$) a ``$+1$" and to each odd lattice site a ``$-1$". Obviously, the operator $\Gamma_\mu$ is diagonal consisting of equally many eigenvalues $\pm1$ and only acts on the lattice part $\widehat{V}$ of the Hilbert space. This artificial operator  satisfies the following commutation relations with the transfer matrices,
\begin{equation}\label{com:gam-trans}
[\Gamma_\mu,T_\mu]_+=0\text{ and }[\Gamma_\mu,T_\nu]_-\overset{\mu\neq\nu}{=}0.
\end{equation}
Combining this artificial operator with the $\gamma$-matrix $\gamma_\mu$, i.e.
\begin{equation}\label{def:gamma5mu}
 \Gamma_\mu^{(5)}=\Gamma_\mu\gamma_\mu
\end{equation}
(note, we do {\bf not} sum over $\mu$),  then we obtain the anti-commutation relation
\begin{equation}\label{com:gam-Dir}
[\Gamma_\mu^{(5)},D]_+=\sum_{\nu=1}^d[\Gamma_\mu\gamma_\mu,(T_\nu-T_\nu^{\dagger})\gamma_\nu]_+=0.
\end{equation}

The lattice may not only have one direction with an even partition. For example for staggered fermions all $L_\mu$ are even. For each direction with an even partition we construct an operator $\Gamma_\mu^{(5)}$ and each of them satisfies~\eqref{com:gam-Dir}. Suppose $N_{\rm L}\leq d$ directions have an even partition and we choose these directions as $\mu=1,\ldots,N_{\rm L}$ without loss of generality. We underline again that the different boundary conditions for spatial and temporal directions play a minor role. Additionally we define the operator $\Gamma_{N_{\rm L}+1}^{(5)}=\gamma^{(5)}$ depending on whether the space-time dimension $d$ is even. The operators $\{\Gamma_{j}^{(5)}\}_{j=1,\ldots,N_{\rm Cl}}$ also build a Clifford algebra of $N_{\rm Cl}=N_{\rm L}+[d+1]_2$ generators, i.e.
\begin{equation}\label{Clifford-big}
 [\Gamma_i^{(5)},\Gamma_j^{(5)}]_+=2\delta_{ij}\eins_{d_{\mathcal{H}}},\ \tr\Gamma_j^{(5)}=0\text{ and }(\Gamma_j^{(5)})^\dagger=\Gamma_j^{(5)}.
\end{equation}
The function $[d+1]_2$ is $1$ if $d$ is even and vanishes for odd $d$. All operators $\{\Gamma_{j}^{(5)}\}_{j=1,\ldots,N_{\rm Cl}}$ anti-commute with the Dirac operator, see eq.~\eqref{com:gam-Dir}.

In the case there is a charge conjugation operator
\begin{equation}\label{def:Clat}
\begin{split}
C_{\rm lat}=&K\zeta\chi\text{ and }\\
C_{\rm lat}^2=&(-1)^{(d+2)(d+1)d(d-1)/8}\sign[(K\zeta)^2]\eins_{d_{\mathcal{H}}},
\end{split}
\end{equation}
namely for real and quaternion representations $r$, with
\begin{equation}\label{com:Clat-Dir}
 [C_{\rm lat},\imath^{d(d-1)/2}D]_-=0,
\end{equation}
we have furthermore the  relation
\begin{equation}\label{com:Clat-gam}
 [C_{\rm lat},\imath^{d(d-1)/2}\Gamma_j^{(5)}]_-=0
\end{equation}
for all $j=1,\ldots,N_{\rm Cl}$.

\section{Symmetry Analysis}\label{sec:symmetry}

The relations~\eqref{com:gam-Dir}, \eqref{Clifford-big}, \eqref{com:Clat-Dir}, and \eqref{com:Clat-gam} together with the definitions~\eqref{def:gamma5mu} and \eqref{def:Clat} completely determine the degeneracy of the eigenvalues and the symmetry class of the lattice Dirac operator $D$ including the symmetry breaking pattern. This has to be done for even and odd $N_{\rm Cl}$, separately. We do this by pursuing the same ideas as in the discussion for the two-dimensional lattice QCD-Dirac operator~\cite{2DstagQCD}.

\subsection{Even number $N_{\rm Cl}$ of Clifford generators}\label{sec:Neven}

When the number $N_{\rm Cl}$ of generators of the Clifford algebra anti-commuting with the Dirac operator is even one can construct an $N_{\rm Cl}+1$'st Hermitian, unitary operator $\hat{\Gamma}^{(5)}=\imath^{N_{\rm Cl}(N_{\rm Cl}-1)/2}\prod_{j=1}^{N_{\rm Cl}}\Gamma_j^{(5)}$ which anti-commutes with all generators but commutes with the naive Dirac operator $D$. Multiplying $D$ with $\hat{\Gamma}^{(5)}$ one obtains the commutation relations
\begin{equation}\label{com:Dgamm-gamm}
[D\hat{\Gamma}^{(5)},\Gamma_j^{(5)}]_-=0\text{ for all }j=1,\ldots,N_{\rm Cl}.
\end{equation}
Since the combinations $\Gamma_i^{(5)}\pm\imath \Gamma_j^{(5)}$ are nil-potent the set $\{\Gamma_{j}^{(5)}\}_{j=1,\ldots,N_{\rm Cl}}$ generates a direct sum of fundamental representations of the Lie algebra of the unitary group $\U(2^{N_{\rm Cl}/2})$ by multiplication of elements and of scalars and by addition. The fundamental representation of the Clifford algebra generated by $N_{\rm Cl}$ Hermitian elements is unique up to unitary transformations because $N_{\rm Cl}$ is even. Thus there is a unitary matrix $U\in\U(d_{\mathcal{H}})$ with
\begin{equation}\label{Gam-unitary}
U\Gamma_j^{(5)}U^{\dagger}=\eins_{d_{\mathcal{H}}/2^{N_{\rm Cl}/2}}\otimes\gamma'_j\text{ for all }j=1,\ldots,N_{\rm Cl},
\end{equation}
with $\{\gamma'_j\}_{j=1,\ldots,N_{\rm Cl}}$ in the fundamental representation of the Clifford algebra in $2^{N_{\rm Cl}/2}$ dimensions. The tensor notation shall emphasize the splitting of the Hilbert space into two terms. We will also employ it in the following as a bookkeeping tool.

Equation~\eqref{Gam-unitary} also implies
\begin{equation}\label{Gam5-unitary}
\begin{split}
U\hat{\Gamma}^{(5)}U^{\dagger}=&\eins_{d_{\mathcal{H}}/2^{N_{\rm Cl}/2}}\otimes{\gamma'}^{(5)}\\
=& \imath^{N_{\rm Cl}(N_{\rm Cl}-1)/2}\eins_{d_{\mathcal{H}}/2^{N_{\rm Cl}/2}}\otimes\prod_{j=1}^{N_{\rm Cl}}\gamma'_j.
\end{split}
\end{equation}
With the aid of Schur's lemma~\cite{Schur}, the commutation relation~\eqref{com:Dgamm-gamm} yields $UD\hat{\Gamma}^{(5)}U^\dagger=D_{\rm red}\otimes\eins_{2^{N_{\rm Cl}/2}}$ or equivalently
\begin{equation}\label{Dlat-Neven}
UDU^\dagger=D_{\rm red}\otimes{\gamma'}^{(5)}
\end{equation}
with a reduced lattice Dirac operator $D_{\rm red}=-D_{\rm red}^\dagger$ only acting on a Hilbert space of dimension $d_{\mathcal{H}}/2^{N_{\rm Cl}/2}$. In section~\ref{sec:MC-sim} we write $D_{\rm red}$ more explicitly by  choosing a basis in the spinor space.

In the case that there is no additional anti-unitary symmetry (complex representation of the gauge group) the eigenvalues of $D$ are $2^{N_{\rm Cl}/2-1}$-fold degenerate. Moreover all eigenvalues come in ``chiral" pairs $(\lambda,-\lambda)$ apart from the case $N_{\rm Cl}=0$. These ``chiral" pairs are reminiscent of the chirality of the original lattice Dirac operator $D$. The Dirac operator will have the global symmetries of the three dimensional continuum theory.

In the case of a real or quaternion representation, we have also to consider the transformation of the charge conjugation operator $C_{\rm lat}$. We define $C'_{\rm lat}=UC_{\rm lat}U^\dagger=K\zeta'\otimes\chi'$ with unitary matrices $\zeta'\in\U(d_{\mathcal{H}}/2^{N_{\rm Cl}/2})$ and $\chi'\in\U(2^{N_{\rm Cl}/2})$. The commutation relations~\eqref{com:Clat-gam} and \eqref{com:Clat-Dir} become
\begin{equation}\label{com:Clat-gam-red}
\begin{split}
 [C'_{\rm lat},\imath^{d(d-1)/2}\eins_{d_{\mathcal{H}}/2^{N_{\rm Cl}/2}}\otimes\gamma'_j]_-=&0\\
 \Rightarrow  [C'_{\rm lat},\imath^{N_{\rm Cl}(N_{\rm Cl}-1)/2}\eins_{d_{\mathcal{H}}/2^{N_{\rm Cl}/2}}\otimes{\gamma'}^{(5)}]_-=&0
 \end{split}
\end{equation}
and
\begin{equation}\label{com:Clat-Dir-red}
 \begin{split}
 &\pm[C'_{\rm lat},\imath^{d(d-1)/2}D_{\rm red}\otimes{\gamma'}^{(5)}]_-\\
 =&[C_{\rm red},\imath^{d_{\rm red}(d_{\rm red}-1)/2}D_{\rm red}]_-\otimes \imath^{N_{\rm Cl}(N_{\rm Cl}-1)/2}\chi'{\gamma'}^{(5)}=0
 \end{split}
\end{equation}
with $C_{\rm red}=K\zeta'$ and $d_{\rm red}=d-N_{\rm L}$, respectively. To derive eq.~\eqref{com:Clat-Dir-red} we need
\begin{equation}
\imath^{d(d-1)/2}=\pm\imath^{d_{\rm red}(d_{\rm red}-1)/2+N_{\rm Cl}(N_{\rm Cl}-1)/2}
\end{equation}
which is true because $N_{\rm L}=N_{\rm Cl}-[d+1]_2$ with $[d+1]_2=0,1$ and $d[d+1]_2$ is always even.

What remains to be calculated is the square $C_{\rm red}^2$. For this purpose we notice that $(C'_{\rm lat})^2=C_{\rm lat}^2$. Furthermore the unitary matrix $\chi'$ can be chosen as
\begin{equation}
\begin{split}
 \chi'=&({\gamma'}^{(5)})^{d_{\rm red}(d_{\rm red}-1)/2}\\
 &\times\left\{\begin{array}{cl} \imath^{m}\prod_{j=1}^{2m}\gamma'_{2j}, & N_{\rm Cl}=4m,\\ \imath^{m+1}\prod_{j=1}^{2m+2}\gamma'_{2j-1}, & N_{\rm Cl}=4m+2, \end{array}\right.
 \end{split}
\end{equation}
because of the commutation relations
\begin{equation}\label{com:Clat-gam-red-b}
[K\chi',\imath^{d(d-1)/2}\gamma'_j]_-=0\text{ for all }j=1,\ldots,N_{\rm Cl}.
\end{equation}
All other choices of $\chi'$ are unitarily equivalent. The commutation relations~\eqref{com:Clat-gam-red-b} directly follow from the first line of eq.~\eqref{com:Clat-gam-red}, since the first component of the tensor product is trivial. Employing the relation $C_{\rm lat}^2=(C'_{\rm lat})^2=C_{\rm red}^2\otimes(K\chi')^2$ we obtain
\begin{equation}\label{Clatred-square-even}
\begin{split}
C_{\rm red}^2=&(-1)^{(d_{\rm red}+2)(d_{\rm red}+1)d_{\rm red}(d_{\rm red}-1)/8}\\
&\times\sign[(K\zeta)^2]\eins_{d_{\mathcal{H}}/2^{N_{\rm Cl}/2}}.
\end{split}
\end{equation}
To simplify the sign we have used the identity
\begin{equation}\label{signs}
\begin{split}
&(-1)^{(d+2)(d+1)d(d-1)/8+(N_{\rm Cl}+2)(N_{\rm Cl}+1)N_{\rm Cl}(N_{\rm Cl}-1)/8}\\
=&(-1)^{(d_{\rm red}+2)(d_{\rm red}+1)d_{\rm red}(d_{\rm red}-1)/8+d_{\rm red}(d_{\rm red}-1)N_{\rm Cl}(N_{\rm Cl}-1)/4}
\end{split}
\end{equation}
which is even true for odd $N_{\rm Cl}$ as can be checked by choosing $d,N_{\rm L}=1,\ldots,8$ because of the periodicity in $8$.

The symmetries of the Dirac operator $D$ experience a shift in the Bott periodic table~\ref{tab:cont-class} due to the lattice directions with even partition. Moreover, the degeneracy of the eigenvalues of the lattice Dirac operator $D$ is either $2^{N_{\rm Cl}/2-1}$-fold or $2^{N_{\rm Cl}/2}$-fold depending on whether $D_{\rm red}$ exhibits Kramer's degeneracy or ``chiral pairs" of eigenvalues or not. We emphasize that the whole Dirac operator $D$ always exhibits ``chiral pairs" of eigenvalues. 

Summarizing the discussion for an even number $N_{\rm Cl}$ of Clifford generators anti-commuting with the Dirac operator, the symmetries of $D$ are shifted via the Bott periodic table~\ref{tab:cont-class} from the original symmetries of the continuum theory in $d$ dimensions to a theory sharing the symmetries of the same continuum theory but only in $d_{\rm red}=d-N_{\rm L}$ dimensions. This is true for all representations. Thus, for any representation the shift of the symmetry class is always to an odd-dimensional continuum theory because $N_{\rm Cl}=N_{\rm L}+[d+1]_2$ is even. This will be different for the case of odd $N_{\rm Cl}$. The symmetry breaking patterns are summarized in table~\ref{tab:lat-class}.

\subsection{Odd number $N_{\rm Cl}$ of Clifford generators}\label{sec:Nodd}

For an odd number $N_{\rm Cl}$ of Clifford generators the corresponding fundamental representation of the Clifford algebra is not unique but there are two. Indeed we have in our case that the product of all generators $\prod_{j=1}^{N_{\rm Cl}}\Gamma_j^{(5)}$ is not proportional to the identity as it would be for the case when it is a multiple of one of the two inequivalent fundamental representations. The product is
\begin{equation}\label{def:tildegam}
\begin{split}
\widetilde{\Gamma}=&\imath^{N_{\rm Cl}(N_{\rm Cl}-1)/2}\prod_{j=1}^{N_{\rm Cl}}\Gamma_j^{(5)}=\Gamma^{(5)}\widetilde{\gamma}\\
=&\imath^{N_{\rm Cl}(N_{\rm Cl}-1)/2}\Gamma^{(5)}\left\{\begin{array}{cc} \left(\prod_{j=1}^{N_{\rm L}}\gamma_j\right)\gamma^{(5)}, & d\in 2\mathbb{N}, \\ \prod_{j=1}^{N_{\rm L}}\gamma_j, & d\in2\mathbb{N}+1, \end{array}\right.
\end{split}
\end{equation}
with $\Gamma^{(5)}=\prod_{j=1}^{N_{\rm L}}\Gamma_j$ a diagonal matrix with an equal number of eigenvalues $\pm1$. Also the matrix $\widetilde{\Gamma}$ has the same number of eigenvalues with $\pm1$ which follows from $\widetilde{\Gamma}=\widetilde{\Gamma}^\dagger=\widetilde{\Gamma}^{-1}$ and $\tr\widetilde{\Gamma}=(\tr \Gamma^{(5)})(\tr\widetilde{\gamma})=0$. Thus we have the same number of both fundamental representations of the Clifford algebra with $N_{\rm Cl}$ elements. Then the matrices $\{\imath\widetilde{\Gamma}\Gamma_j^{(5)}\}_{j=1,\ldots,N_{\rm Cl}}$ build a multiple of only one of the two fundamental representations. In particular we can find a unitary matrix $U\in\U(d_{\mathcal{H}})$ with
\begin{equation}\label{Gam-unitary.b}
\begin{split}
U\Gamma_j^{(5)}U^{\dagger}=&\gamma_{\rm red}^{(5)}\otimes\gamma'_j\text{ for all }j=1,\ldots,N_{\rm Cl},\\
U\widetilde{\Gamma}U^\dagger=&\gamma_{\rm red}^{(5)}\otimes\eins_{2^{(N_{\rm Cl}-1)/2}},
\end{split}
\end{equation}
where $\gamma'_{N_{\rm Cl}}=\imath^{N_{\rm Cl}(N_{\rm Cl}-1)/2}\prod_{j=1}^{N_{\rm Cl}-1}\gamma'_j$ and $\gamma_{\rm red}^{(5)}$ are diagonal matrices with  eigenvalues $\pm1$ each with multiplicity $d_{\mathcal{H}}/2^{(N_{\rm Cl}+1)/2}$. Again the tensor notation shall help to separate the Hilbert space into a space where the naive Dirac operator acts trivially and a reduced Hilbert space.

In the next step we consider the Clifford algebra of $N_{\rm Cl}-1$ elements $\{\imath\eins_{d_{\mathcal{H}}/2^{(N_{\rm Cl}-1)/2}}\otimes\gamma'_{N_{\rm Cl}}\gamma'_j\}_{j=1,\ldots,N_{\rm Cl}-1}$ which commutes with the lattice Dirac operator $UDU^\dagger$, i.e.
\begin{equation}\label{com:Dgamm-gamm-odd}
[UDU^\dagger,\imath\eins_{d_{\mathcal{H}}/2^{(N_{\rm Cl}-1)/2}}\otimes\gamma'_{N_{\rm Cl}}\gamma'_j]_-=0
\end{equation}
 because $[D,\Gamma_{N_{\rm Cl}}^{(5)}\Gamma_j^{(5)}]_-=0$ for all $j=1,\ldots,N_{\rm Cl}$.
Schur's lemma~\cite{Schur} (now the commutation of $D$ with a fundamental representation of $\U(2^{(N_{\rm Cl}-1)/2})$) tells us that the Dirac operator has the form
\begin{equation}\label{Dlat-Nodd}
\begin{split}
UDU^\dagger=&D_{\rm red}\otimes\eins_{2^{(N_{\rm Cl}-1)/2}}\text{ with}\\
D_{\rm red}=&-D_{\rm red}^\dagger\text{ and }[D_{\rm red},\gamma_{\rm red}^{(5)}]_+=0.
\end{split}
\end{equation}
The last equality follows from the anti-commutation relation of $[D,\widetilde{\Gamma}]_+=0$. An explicit form of $D_{\rm red}$ is given in section~\ref{sec:MC-sim}.

For complex representations the discussion above implies that all eigenvalues are $2^{(N_{\rm Cl}-1)/2}$-fold degenerate. Furthermore the reduced Dirac operator $D_{\rm red}$ has a chiral form with no further symmetries. Thus, the lattice theory should share the symmetries of the continuum theory in four dimensions.

When we consider a real or a quaternion representation $r(\mathfrak{g})$ of a gauge group, we have to evaluate the implications to the anti-unitary symmetries. We consider the anti-unitary operator in the new basis $C'_{\rm lat}=UC_{\rm lat}\widetilde{\Gamma}^{N_{\rm Cl}(N_{\rm Cl}-1)/2}U^\dagger=K\zeta'\otimes\chi'$ which is equivalent to the original anti-unitary operator $C_{\rm lat}$. The factor $\widetilde{\Gamma}^{N_{\rm Cl}(N_{\rm Cl}-1)/2}$ is introduced in the new anti-unitary operator because of the following anti-commutation relation with the reduced Dirac operator,
\begin{equation}\label{com:Clat-Dir-red-odd}
 \begin{split}
 &U[C_{\rm lat}\widetilde{\Gamma}^{N_{\rm Cl}(N_{\rm Cl}-1)/2},\imath^{d(d-1)/2+N_{\rm Cl}(N_{\rm Cl}-1)/2}D]_-U^\dagger\\
 =&[C'_{\rm lat},\imath^{d(d-1)/2+N_{\rm Cl}(N_{\rm Cl}-1)/2}D_{\rm red}\otimes\eins_{2^{(N_{\rm Cl}-1)/2}}]_-\\
 =&\pm[C_{\rm red},\imath^{d_{\rm red}(d_{\rm red}-1)/2}D_{\rm red}]_-\otimes\chi'=0,
 \end{split}
\end{equation}
due to the anti-commutation relation $[\widetilde{\Gamma},D]_+=0$ and eq.~\eqref{com:Clat-Dir}.  Again we have chosen the notation $C_{\rm red}=K\zeta'$ and $d_{\rm red}=d-N_{\rm L}$.

Now we need to calculate $C_{\rm red}^2$. For this purpose we need the commutation relations between $C'_{\rm lat}$ and $U\Gamma_j^{(5)}U^\dagger$ which read
\begin{equation}\label{com:Clat-Gamm-odd}
\begin{split}
&[C'_{\rm lat},\imath^{d(d-1)/2}\gamma_{\rm red}^{(5)}\otimes\gamma'_j]_-\\
=&U[C_{\rm lat}\widetilde{\Gamma}^{N_{\rm Cl}(N_{\rm Cl}-1)/2},\imath^{d(d-1)/2}\Gamma_j^{(5)}]_-U^\dagger=0,
\end{split}
\end{equation}
for all $j=1,\ldots,N_{\rm Cl}$. The commutation relation with $U\widetilde{\Gamma}U^\dagger$ is
\begin{equation}\label{com:Clat-Gammtilde-odd}
\begin{split}
&[C'_{\rm lat},\imath^{d_{\rm red}(d_{\rm red}-1)/2}\gamma_{\rm red}^{(5)}\otimes\eins_{2^{(N_{\rm Cl}-1)/2}}]_-\\
=&\pm U[C_{\rm lat}\widetilde{\Gamma}^{N_{\rm Cl}(N_{\rm Cl}-1)/2},\imath^{d(d-1)/2}\prod_{j=1}^{N_{\rm Cl}}\Gamma_j^{(5)}]_-U^\dagger=0
\end{split}
\end{equation}
up to an overall sign. Combining Eqs.~\eqref{com:Clat-Gamm-odd} and \eqref{com:Clat-Gammtilde-odd} yields
\begin{equation}\label{com:Clat-Gamm-odd.b}
[C'_{\rm lat},\imath^{N_{\rm Cl}(N_{\rm Cl}-1)/2}\eins_{d_{\mathcal{H}}/2^{(N_{\rm Cl}-1)/2}}\otimes\gamma'_j]_-=0.
\end{equation}
Thus we can choose
\begin{equation}
\begin{split}
 \chi'=&\left\{\begin{array}{cl} \imath^{m}\prod_{j=1}^{2m}\gamma'_{2j}, & N_{\rm Cl}=4m+1,\\ \imath^{m+1}\prod_{j=1}^{2m+2}\gamma'_{2j-1}, & N_{\rm Cl}=4m+3 \end{array}\right.
 \end{split}
\end{equation}
because all other choices would be unitarily equivalent.

In the last step we employ the relation $(C_{\rm lat}\widetilde{\Gamma}^{N_{\rm Cl}(N_{\rm Cl}-1)/2})^2=C_{\rm red}^2\otimes(K\chi')^2$ which yields us the sign
\begin{equation}\label{Clatred-square-odd}
\begin{split}
C_{\rm red}^2=&(-1)^{(d_{\rm red}+2)(d_{\rm red}+1)d_{\rm red}(d_{\rm red}-1)/8}\\
&\times\sign[(K\zeta)^2]\eins_{d_{\mathcal{H}}/2^{(N_{\rm Cl}-1)/2}}
\end{split}
\end{equation}
because $\sign[(K\chi')^2]=(-1)^{(N_{\rm Cl}+2)(N_{\rm Cl}+1)N_{\rm Cl}(N_{\rm Cl}-1)/8}$ and $\sign[(C_{\rm lat}\widetilde{\Gamma})^2]=(-1)^{d_{\rm red}(d_{\rm red}-1)/2}\sign[C_{\rm lat}^2]$. Moreover we used the identity~\eqref{signs} which is also true for odd $N_{\rm Cl}$.

Combining eqs.~\eqref{Dlat-Nodd}, \eqref{com:Clat-Dir-red-odd} and \eqref{Clatred-square-odd} we can summarize that the eigenvalues of  $D$ are either $2^{(N_{\rm Cl}-1)/2}$ degenerate or   $2^{(N_{\rm Cl}+1)/2}$ degenerate if Kramers degeneracy applies. Moreover the symmetries of $D_{\rm red}$ or equivalently of $D$ are those of the continuum theory at even dimension $d_{\rm red}=d-N_{\rm L}$ and not of dimension $d$. Hence, also for an odd number of Clifford generators $N_{\rm Cl}$ anti-commuting with  the lattice Dirac operator $D$ the symmetries are shifted along the Bott periodic table~\ref{tab:cont-class}. Interestingly, the shift is exactly the same as for even $N_{\rm Cl}$.

\section{Symmetry Breaking Pattern and Zero Modes}\label{sec:sympat}

In the previous sections we have seen that the lattice Dirac operator $D$ may drastically degenerate when some or even all (case of staggered fermions~\cite{staggered}) lattice directions exhibit an even partition of lattice sites. The reduced lattice Dirac operator $D_{\rm red}$ acts on a Hilbert space of dimension $d_{\mathcal{H}}/d_{\rm tri}$ with $d_{\rm tri}=2^{\lfloor N_{\rm Cl}/2\rfloor}$ whose value depends on the fact whether $N_{\rm Cl}$ is even or not. The characteristic polynomial of the lattice Dirac operator with a quark mass $m$ is then
\begin{equation}\label{charpol:NCLeven}
\begin{split}
\det(D+m\eins_{d_{\mathcal{H}}})=&{\det}(D_{\rm red}+m\eins_{d_{\mathcal{H}}/d_{\rm tri}})^{d_{\rm tri}/2}\\
&\times{\det}(-D_{\rm red}+m\eins_{d_{\mathcal{H}}/d_{\rm tri}})^{d_{\rm tri}/2}
\end{split}
\end{equation}
for an even number $N_{\rm Cl}$ of Clifford elements anti-commuting with $D$ and
\begin{equation}\label{charpol:NCLodd}
\det(D+m\eins_{d_{\mathcal{H}}})={\det}(D_{\rm red}+m\eins_{d_{\mathcal{H}}/d_{\rm tri}})^{d_{\rm tri}}
\end{equation}
for odd $N_{\rm Cl}$. Thus the number of physical flavors is enhanced by $d_{\rm tri}$. In particular the symmetry breaking patterns are those of the continuum theory of dimension $d_{\rm red}=d-N_{\rm L}$ with $d_{\rm tri}N_{\rm f}$ flavors. This is also shown in table~\ref{tab:lat-class}. 

The zero modes of the naive lattice Dirac operator are also enhanced by the factor $d_{\rm tri}$. However exact zero modes are only present when $D$ or equivalently $D_{\rm red}$ are in the symmetry class D and DIII because the off-diagonal operators are always of square form (we have always the same number of vectors with positive and negative chirality). The symmetry class D means that $D_{\rm red}$ is a real anti-symmetric matrix of odd dimension with no additional symmetries such that it has one exact zero mode. This zero mode is currently interpreted as a Majorana fermion in condensed matter physics, see \cite{Majorana}. The class DIII implies that $D_{\rm red}$ has a chiral structure whose off-diagonal block is anti-symmetric and odd dimensional such that $D_{\rm red}$ has two zero modes; one has positive chirality and the other one has negative.

We recognize two things. First a QCD-like theory with a complex gauge group representation will never yield a naive Dirac operator with zero modes. Hence we can restrict the discussion about the zero modes on real and quaternion representations.  Second, it strongly depends on the dimension $d_{\mathcal{H}}/d_{\rm tri}$ of the reduced Hilbert space as well as on $d_{\rm red}=d-N_{\rm L}$, the effective dimension reflected by the symmetries of $D_{\rm red}$, whether there is a zero mode or not. Therefore we have to go through the four cases $d_{\rm red}=8\mathbb{N}_0+j$ with $j=1,3,5,7$ where these zero modes may appear.

First we consider $d_{\rm red}\in8\mathbb{N}_0+1$ and a real representation of the gauge group or $d_{\rm red}\in8\mathbb{N}_0+5$ and a quaternion representation. Then we find a symmetry class D when $N_{\rm Cl}$ is even, because this symmetry class does not satisfy a chiral symmetry, and when $d_{\mathcal{H}}/d_{\rm tri}=2^{\lfloor d/2\rfloor}d_rV/2^{N_{\rm Cl}/2}$ is odd where $d_r=2^{c_1}c_2$, with $c_1\in\mathbb{N}_0$ and $c_2$ an odd integer, is the dimension of the representation. We have $N_{\rm L}=N_{\rm Cl}-[d+1]_2\leq d$ and $V=2^{N_{\rm L}+b_1}b_2\in 2^{N_{\rm L}}\mathbb{N}$ with $b_1\in\mathbb{N}_0$ and $b_2$ an odd integer. Then the equation
\begin{equation}
\begin{split}
0=&\left\lfloor\frac{d}{2}\right\rfloor+N_{\rm L}+c_1+b_1-\frac{N_{\rm L}+[d+1]_2}{2}\\
=&\frac{d+N_{\rm L}-1}{2}+c_1+b_1
\end{split}
\end{equation}
has to be satisfied to find a zero mode. However this will only be the case for $d=d_{\rm red}=1$ and $N_{\rm L}=b_1=c_1=0$. Since we excluded the lattice theory in one dimension because  there is no spontaneous symmetry breaking, we conclude that the symmetry class D never shows up for the naive discretization. In other words, when $D$ and, thus, $D_{\rm red}$ are real anti-symmetric matrices they will be always of even dimension (Cartan class B) regardless what QCD-like lattice gauge theory one considers and which effective dimension $d_{\rm red}=d-N_{\rm L}$ we consider.

In the third and fourth case we consider $d_{\rm red}\in8\mathbb{N}_0+2$ and a real representation of the gauge group or $d_{\rm red}\in8\mathbb{N}_0+6$ and a quaternion representation. Then, $N_{\rm Cl}$ has to be odd because the symmetry class DIII exhibits a chiral structure. For a zero mode the off-diagonal block has to be odd dimensional, i.e. $d_{\mathcal{H}}/(2d_{\rm tri})=2^{\lfloor d/2\rfloor}d_rV/2^{(N_{\rm Cl}+1)/2}$ has to be odd. The additional division by to $2$ results from the chiral structure of the matrix $D_{\rm red}$. We have to solve the equation
\begin{equation}
\begin{split}
0=&\left\lfloor\frac{d}{2}\right\rfloor+N_{\rm L}+c_1+b_1-\frac{N_{\rm L}+[d+1]_2+1}{2}\\
=&\frac{d+N_{\rm L}-2}{2}+c_1+b_1.
\end{split}
\end{equation}
This is only satisfied when $d=d_{\rm red}=2$ and $N_{\rm L}=b_1=c_1=0$ because we exclude the one-dimensional case. Indeed this case was found in the simulations performed in \cite{2DstagQCD}. Apart from this particular case again no generic zero modes will be found for the naive lattice Dirac operator.

Let summarize the discussion about the zero modes. Excluding one- and two-dimensional theories, all naive lattice Dirac operators will never show generic zero modes independently of the representation of the gauge group, of the space-time dimension and of the partition of the lattice.

We summarize the results above in table~\ref{tab:lat-class}.

\section{Explicit Representation of the Dirac Operator}\label{sec:MC-sim}

At last we want to derive an explicit representation of the reduced lattice Dirac operator $D_{\rm red}$ which is the staggered Dirac operator for $N_{\rm L}=d$. To achieve this we define the $d_{\mathcal{H}}\times d_{\mathcal{H}}$ unitary matrices
\begin{equation}\label{def:V}
\begin{split}
V^{(1)}_j=&\frac{1}{2}\left(\eins_{d_{\mathcal{H}}}+\Gamma_j+\gamma_j-\Gamma_j\gamma_j\right),\\
V^{(2)}_j=&\frac{1}{2}\left(\eins_{d_{\mathcal{H}}}+\Gamma_j-\gamma_j+\Gamma_j\gamma_j\right),\\
V^{(3)}_j=&\frac{1}{2}\left(\eins_{d_{\mathcal{H}}}-\Gamma_j+\gamma_j+\Gamma_j\gamma_j\right)
\end{split}
\end{equation}
for $j=1,\ldots,N_{\rm L}$. These matrices are Hermitian, $V^{(l)}_j={V^{(l)}_j}^\dagger$, and, thus, self-inverse, $V^{(l)}_j{V^{(l)}_j}^\dagger=\eins_{d_{\mathcal{H}}}$. Moreover they satisfy
\begin{equation}\label{VpVm}
V^{(1)}_jV^{(2)}_j=\Gamma_j,\ V^{(1)}_jV^{(3)}_j=\gamma_j\text{ and }V^{(2)}_jV^{(3)}_j=\Gamma_j\gamma_j.
\end{equation}
The product of the matrices $V^{(1)}_j$, i.e.
\begin{equation}\label{def:V-prod}
\tilde{U}=V^{(1)}_1V^{(1)}_2\cdots V^{(1)}_{N_{\rm L}},
\end{equation}
 will serve as the change of basis we are looking for to identify  $D_{\rm red}$. We want to emphasize that $\tilde{U}$ is not necessarily equal to the unitary matrix $U$ from the previous subsections. Especially the charge conjugation operator $C_{\rm lat}$ will be different from $C'_{\rm lat}$ after conjugation with $\tilde{U}$ though it will be equivalent. Our particular choice for $\tilde{U}$ is that the transformed  Dirac operator $\tilde{U}^\dagger D\tilde{U}$ will have a simple form where $D_{\rm red}$ can be readily read off.
 
 For the transformation of the Dirac operator we need the commutation relations
 \begin{equation}\label{com:DgV}
 \begin{split}
  (T_i-T_i^\dagger)\gamma_i V_i^{(1)}=&V_i^{(3)}(T_i-T_i^\dagger)\gamma_i,\\
  (T_i-T_i^\dagger)\gamma_i V_j^{(1)}\overset{i\neq j}{=}&V_j^{(2)}(T_i-T_i^\dagger)\gamma_i,\\
  \gamma_i V_j^{(2)}\overset{i\neq j}{=}&V_j^{(1)}\gamma_i.
  \end{split}
 \end{equation}
 Again we want to recall that we do not use Einstein's summation convention. Combining the commutation relations~\eqref{com:DgV} with Eqs.~\eqref{def:lat-Dirac}, \eqref{VpVm}, and \eqref{def:V-prod} the Dirac operator in the new basis looks as follows
 \begin{equation}\label{Dirac-new-basis}
 \begin{split}
 \tilde{U}^\dagger D\tilde{U}=&\sum_{\mu=1}^dV^{(1)}_{N_{\rm L}}\cdots V^{(1)}_{1}V^{(2)}_{1}\cdots V^{(2)}_{\mu-1}V^{(3)}_{\mu}V^{(2)}_{\mu+1}\cdots V^{(2)}_{N_{\rm L}}\\
 &\times(T_\mu-T_\mu^\dagger)\gamma_\mu\\
 =&\sum_{\mu=1}^{N_{\rm L}} \left(\prod_{j=1}^{\mu-1}\Gamma_j\right)(T_\mu-T_\mu^\dagger)\\
 &+\sum_{\mu=N_{\rm L}+1}^d \left(\prod_{j=1}^{N_{\rm L}}\Gamma_j\right)(T_\mu-T_\mu^\dagger)\gamma_\mu.
 \end{split}
 \end{equation}
 We notice that the covariant derivatives in the first $N_{\rm L}$ directions act trivially in the spinor space. Hence, we may choose a basis of the generalized $\gamma$-matrices for $\mu>N_{\rm L}$ in the following way,
 \begin{equation}\label{new-gamma-cl-odd}
 \begin{split}
 \gamma_\mu=\tilde{\gamma}_\mu\otimes\eins_{2^{(N_{\rm Cl}-1)/2}}
 \end{split}
 \end{equation}
 for $N_{\rm Cl}$ being odd and
 \begin{equation}\label{new-gamma-cl-even}
 \begin{split}
 \gamma_\mu=\tilde{\gamma}_\mu\otimes\sigma_3^{\otimes N_{\rm Cl}/2}
 \end{split}
 \end{equation}
 for even $N_{\rm Cl}$. The matrix $\sigma_3^{\otimes j}$ is the tensor product of the third Pauli matrix taken $j$-times. The matrices $\tilde{\gamma}_\mu$ build the generalized $\gamma$-matrices in $d_{\rm red}=d-N_{\rm L}$ dimensions. The remaining $\gamma$-matrices $\gamma_\mu$ with $\mu\leq N_{\rm L}$ are then of the form $\sigma_3^{\otimes J}\otimes \gamma'_\mu$ for odd $N_{\rm Cl}$ and $\eins_{2^{\otimes J}}\otimes \gamma'_\mu$ for even $N_{\rm Cl}$ with $J=\lfloor d/2\rfloor-\lfloor N_{\rm Cl}/2\rfloor$, but this is not important anymore. Equation~\eqref{Dirac-new-basis} is enough to read off the reduced lattice Dirac operator which is
 \begin{equation}\label{Dirac-red} 
 D_{\rm red}=\sum_{\mu=1}^{N_{\rm L}} D^{\rm (red)}_\mu+\sum_{\mu=N_{\rm L}+1}^d D^{\rm (red)}_\mu\gamma_\mu
 \end{equation}
 with the new covariant derivatives
 \begin{equation}\label{new-cov-der-NLs}
 \begin{split}
  D^{\rm (red)}_\mu|\psi(x)\rangle=&(-1)^{\sum_{j=1}^{\mu-1}x_j}\left((-1)^{\delta_{jd}\delta_{x_dL_d}}U_\mu(x)|\psi(x+e_\mu)\rangle\right.\\
  &\left.-(-1)^{\delta_{jd}\delta_{x_dL_1}}U_\mu^\dagger(x)|\psi(x-e_\mu)\rangle\right)
 \end{split}
 \end{equation}
 for $\mu\leq N_{\rm L}$ and
 \begin{equation}\label{new-cov-der-NLl}
 \begin{split}
  D^{\rm (red)}_\mu|\psi(x)\rangle=&(-1)^{\sum_{j=1}^{N_{\rm L}}x_j}\left((-1)^{\delta_{jd}\delta_{x_dL_d}}U_\mu(x)|\psi(x+e_\mu)\rangle\right.\\
  &\left.-(-1)^{\delta_{jd}\delta_{x_dL_1}}U_\mu^\dagger(x)|\psi(x-e_\mu)\rangle\right)
 \end{split}
 \end{equation}
 for $\mu> N_{\rm L}$. We underline that the only difference between Eqs.~\eqref{new-cov-der-NLs} and \eqref{new-cov-der-NLl} is the overall sign.
 
 For $N_{\rm L}=d$ the Dirac operator~\eqref{Dirac-red} automatically reduces to the staggered Dirac operator~\cite{staggered}. In the case that $-U(x)$ is in the considered representation of the gauge group when $U(x)$ is, the signs can be absorbed. For example this is the case for the fundamental representation of ${\rm SU}(2)$.
 
For the formulas~\eqref{Dirac-red}, \eqref{new-cov-der-NLs} and \eqref{new-cov-der-NLl} we assumed that the temporal direction has an odd partition of lattice sites. In the case that the temporal direction has an even number of lattice sites we switch in the formulas the directions $\mu=1$ and $\mu=d$.

In figures~\ref{fig1} and \ref{fig2} we compare Monte Carlo simulations of quenched QCD lattice Dirac operators in the naive discretization and in the strong coupling limit ($\beta\to\infty$, group elements are drawn from the Haar measure) with random matrix theory results. We have chosen several lattices, dimensions and representations of the gauge groups ${\rm SU}(2)$ and ${\rm SU}(3)$. The agreement with the analytical random matrix results recalled in appendix~\ref{sec:RMT} is quite good according to the very small sizes of the lattices.

\section{Conclusions}\label{sec:conclusio}

One well-known consequence of the naive discretization of the Dirac operator is the change of its spectral properties, see \cite{4DstagQCD.a,4DstagQCD.b,3DstagQCD.a,3DstagQCD.b,2DstagQCD}. We analyzed this change for an arbitrary QCD-like gauge theory and at arbitrary space-time dimension $d\geq2$. When one or more directions have an even number of lattice sites the degeneracy increases exponentially. This also results into an increase of the number of flavors (neglecting the doublers still comprised in the continuum limit) by $N_{\rm f}\rightarrow d_{\rm tri}N_{\rm f}$ with $d_{\rm tri}=2^{\lfloor (N_{\rm L}+[d+1]_2)/2\rfloor}$ and $N_{\rm L}$ the number of directions with even parity.

For $d>2$ the Dirac operator has no zero modes such that the topological charge $\nu$ will be zero. Hence all zero modes can only appear when taking the continuum limit. For example Follana et al.~\cite{Follana} have analyzed how these zero modes show up for staggered lattice configurations which converge to configurations with non-trivial topological charge. To decide whether a mode is a ``would-be" zero mode or not was decided by them via measuring the chirality of the individual modes. We have not considered this particular issue here.

Moreover the anti-unitary symmetries and chiral symmetries experience a shift along the symmetry classification of the continuum symmetries, see table~\ref{tab:cont-class} as well as \cite{contclass}. This shift is explicitly given by $d\rightarrow d_{\rm red}=d-N_{\rm L}$. Thus the symmetries of the staggered Dirac operator~\cite{staggered} are always the one of the corresponding continuum theories at dimension $d=8$ regardless what original dimension was chosen as long as the dimension is $d\geq2$. In particular the symmetry breaking patterns will be those at $d=8$.

Additionally we derived an explicit form of the non-degenerate part of the Dirac operator. It reduces to the staggered Dirac operator~\cite{staggered} when all directions have an even partition. We performed Monte Carlo simulations with this reduced Dirac operator in the strong coupling limit and in the three-, four- and five-dimensional quenched theory. The spectral statistics of the lowest eigenvalues were compared with random matrix theory results. The good agreement of the numerics with the analytical results confirm our predictions. This agreement is at least as good as it was found in four dimensions for the staggered fermions~\cite{4DstagQCD.a} despite the fact that the lattices we simulated are very small.

Our results may yield a basis to understand the continuum limit of staggered and naive fermions. In particular it is still an unsolved problem whether the global symmetries change to those of the correct continuum theory. Here we have to say that the weak coupling limit studied in~\cite{4DstagQCD.b} nurtures some doubt because on the smallest scales the global symmetries of the lattice Dirac operator always shows up. When assuming, nonetheless, that the continuum limit exists, one can study the infra-red spectrum of the Dirac operator and, hence, the lightest pseudo scalar mesons with random matrix theory. First attempts in this direction were already done in four~\cite{Osborn} and three~\cite{3DstagQCD.b} dimensions.

\vspace*{-0.4cm}
\section*{Acknowledgements}

We thank Jacobus Verbaarschot for fruitful discussions and  Gernot Akemann for helpful comments on the first draft. MK acknowledges support by the grant AK35/2-1 ``Products of Random Matrices" of the German research council (DFG).

\appendix

\section{Classes of Hermitian Random Matrices}\label{sec:RMT}

There are ten Hermitian symmetric matrix spaces~\cite{Leveldensity,tenfoldway}. Five symmetry classes satisfy a chiral symmetry and the other five do not. Eight of the ten classes have an anti-unitary symmetry with respect to the complex conjugation. The corresponding anti-unitary operator squares to $+\eins$ and $-\eins$ each for four classes. The ten classes are listed in table~\ref{tab:RMT} with the corresponding Cartan classification label, see \cite{threefoldway,tenfoldway}.

The probability density can be chosen Gaussian for all ten classes, i.e.
\begin{equation}\label{def:Prob}
 P(H)\propto\exp\left[-\frac{\tr H^2}{2\sigma^2}\right],
\end{equation}
with variance $\sigma$ which may vary from one class to another. The variance can be read off from the joint probability densities~\eqref{jpdf-a} and \eqref{jpdf-b}. The random matrix $\imath H$ exhibits the spectral statistics of the lowest eigenvalues of QCD-like Dirac operators satisfying the same unitary and anti-unitary symmetries as $\imath H$.

The joint probability density of the eigenvalues $\lambda$ of the random matrix $H$ for the classes A (Hermitian matrices), AI (real symmetric matrices) and AII (Hermitian self-dual matrices) can be summarized in one formula
\begin{equation}\label{jpdf-a}
p(\lambda)\propto|\Delta_n(\lambda)|^{\beta_{\rm D}}\prod_{j=1}^n\exp\left[-\frac{\beta_{\rm D} }{2n}\lambda_j^2\right],
\end{equation}
while the other seven Hermitian matrix ensembles follow the formula
\begin{equation}\label{jpdf-b}
p(\lambda)\propto|\Delta_n(\lambda^2)|^{\beta_{\rm D}}\prod_{j=1}^n\lambda_j^{\alpha_{\rm D}}\exp\left[-\frac{\beta_{\rm D} }{2n}\lambda_j^2\right].
\end{equation}
In the latter case the eigenvalues of the matrices come in ``chiral pairs" $(\lambda_j,-\lambda_j)$ though not all these matrices satisfy a chiral symmetry, e.g. the imaginary anti-symmetric matrices are not chiral but its eigenvalues appear in ``chiral pairs". We recall the Vandermonde determinant
\begin{equation}\label{Vandermonde}
 \Delta_n(\lambda)=\prod_{1\leq a<b\leq n}(\lambda_b-\lambda_a)=\det[\lambda_a^{b-1}]_{a,b=1,\ldots,n}.
\end{equation}
The index $\beta_{\rm D}$ is the Dyson index and determines the strength of the level repulsion between the eigenvalues. The parameter $\alpha_{\rm D}$ is related to the topological charge and is the origin for the level repulsion from the origin.

One important spectral quantity is the microscopic level density. It is a constant for those three ensembles which do not exhibit ``chiral pairs" of eigenvalues, namely real symmetric, Hermitian and Hermitian self-dual matrices. For the other seven ensembles the quenched microscopic level density is non-trivial and has the form~\cite{Leveldensity}
\begin{equation}\label{level-dens}
\begin{split}
\rho_{\nu_{\rm D}}^{(\beta_{\rm D}=1)}(x)=&\frac{|x|}{2}\left(J_{\nu_{\rm D}}^2(x)-J_{\nu_{\rm D}+1}(x)J_{\nu_{\rm D}-1}(x)\right)\\
&+\frac{1}{2}J_{\nu_{\rm D}}(|x|)\left(1-\int_0^{|x|}J_{\nu_{\rm D}}(x')dx'\right),\\
\rho_{\nu_{\rm D}}^{(\beta_{\rm D}=2)}(x)=&\frac{|x|}{2}\left(J_{\nu_{\rm D}}^2(x)-J_{\nu_{\rm D}+1}(x)J_{\nu_{\rm D}-1}(x)\right),\\
\rho_{\nu_{\rm D}}^{(\beta_{\rm D}=4)}(x)=&|x|\left(J_{2\nu_{\rm D}}^2(2x)-J_{2\nu_{\rm D}+1}(2x)J_{2\nu_{\rm D}-1}(2x)\right)\\
&-J_{2\nu_{\rm D}}(2|x|)\left(\frac{1}{2}-\int_{|x|}^\infty J_{2\nu_{\rm D}}(2x')dx'\right).
\end{split}
\end{equation}
The function $J_\nu(x)=\int_{-\pi}^\pi \exp[\imath x\sin(\varphi)-\imath\nu\varphi]d\varphi/(2\pi)$ is the Bessel function of the first kind.
The densities are normalized such that $\lim_{|x|\to\infty}\rho(x)=1/\pi$. The index $\nu_{\rm D}$ is related to $\alpha_{\rm D}$ and the topological charge $\nu$ and can be read off from table~\ref{tab:RMT}.

The densities~\eqref{level-dens} together with the degree of degeneracy of the eigenvalues and the number of the generic zero modes are ideal for deciding to which of the seven symmetry classes a specific spectrum belongs. But what is with the classes of real symmetric, Hermitian and Hermitian self-dual matrices? For these three classes another quantity is needed which is the level spacing distribution. It describes the distribution of the spacing between adjacent eigenvalues. This distribution is very well described by Wigner's surmise~\cite{Wigner}
\begin{equation}\label{Wigner}
\begin{split}
p_{\rm sp}(s)=&2\frac{(\Gamma[(\beta_{\rm D}+2)/2])^{\beta_{\rm D}+1}}{(\Gamma[(\beta_{\rm D}+1)/2])^{\beta_{\rm D}+2}}s^{\beta_{\rm D}}\\
&\times\exp\left[-\left(\frac{\Gamma[(\beta_{\rm D}+2)/2]}{\Gamma[(\beta_{\rm D}+1)/2]}\right)^2s^2\right]
\end{split}
\end{equation}
with the Gamma function $\Gamma(x)$. A better approximation of the level spacing distribution is via a Pad\'e expansion which converges rapidly to the true level spacing distribution, see \cite{Dietz}. Though the distribution~\eqref{Wigner} is not the exact result of the level spacing distribution for $n\to\infty$ it is a good approximation. Its root-mean-square deviation to the correct expression is much less than one per mill.

In the Monte Carlo simulations shown in figures~\ref{fig1} and \ref{fig2}, we make use of eqs.~\eqref{level-dens} and \eqref{Wigner}. They are the analytical curves we compare with the numerics. In this way we confirm our predictions.

\begin{widetext}\
\begin{table} 
\begin{tabular}{|c||c|c|c|}
  \hline
  \diagbox{\ $d$}{\hspace*{-0.25cm}Theory}
                 & real representation & complex representation & quaternion representation  \\
  \hline\hline
  $8m$			
  &	$\overset{\ }{\begin{array}{c} [\gamma^{(5)},\mathcal{D}]_+=[\mathcal{C},\mathcal{D}]_-=[\mathcal{C},\gamma^{(5)}]_-=0,\\ \mathcal{C}^2= \eins \\ \U(2N_{\rm f})\rightarrow\USp(2N_{\rm f}) \\ \text{B$\mid$DI},\ \chi{\rm GOE}_\nu(n),\ \nu\in\mathbb{Z}\end{array}}$		&	$\overset{\ }{\begin{array}{c} {[\gamma^{(5)},\mathcal{D}]_+=0} \\ \U(N_{\rm f})\times\U(N_{\rm f})\rightarrow\U(N_{\rm f}) \\ \text{AIII},\ \chi{\rm GUE}_\nu(n),\ \nu\in\mathbb{Z}\end{array}}$		&	$\overset{\ }{\begin{array}{c} [\gamma^{(5)},\mathcal{D}]_+=[\mathcal{C},\mathcal{D}]_-=[\mathcal{C},\gamma^{(5)}]_-=0,\\ \mathcal{C}^2= -\eins \\ \U(2N_{\rm f})\rightarrow\Ort(2N_{\rm f}) \\ \text{CII},\ \chi{\rm GSE}_\nu(n),\ \nu\in\mathbb{Z} \end{array}}$			\\ \hline
  $8m+1$		
  &	$\overset{\ }{\begin{array}{c} [\mathcal{C},\mathcal{D}]_-=0,\ \mathcal{C}^2= \eins \\ \Ort(2N_{\rm f})\rightarrow\U(N_{\rm f}) \\ \text{B$\mid$D},\ {\rm GAOE}_\nu(n),\ \nu\in\mathbb{Z}_2\end{array}}$	 &	$\overset{\ }{\begin{array}{c} {\rm no\ further\ symmetries} \\ \U(2N_{\rm f})\rightarrow\U(N_{\rm f})\times\U(N_{\rm f})\\ \text{A},\ {\rm GUE}(n)\end{array}}$		&	$\overset{\ }{\begin{array}{c} [\mathcal{C},\mathcal{D}]_-=0,\ \mathcal{C}^2= -\eins \\ \USp(2N_{\rm f})\rightarrow\U(N_{\rm f}) \\ \text{C},\ {\rm GASE}(n) \end{array}}$	\\ \hline  
  $8m+2$		
  &	$\overset{\ }{\begin{array}{c} [\gamma^{(5)},\mathcal{D}]_+=[\mathcal{C},\mathcal{D}]_+=[\mathcal{C},\gamma^{(5)}]_+=0,\\ \mathcal{C}^2= -\eins \\ \Ort(2N_{\rm f})\times\Ort(2N_{\rm f})\rightarrow\Ort(2N_{\rm f}) \\ \text{DIII},\ {\rm GBSE}_\nu(n),\ \nu\in\mathbb{Z}_2\end{array}}$	&	$\overset{\ }{\begin{array}{c} {[\gamma^{(5)},\mathcal{D}]_+=0} \\ \U(N_{\rm f})\times\U(N_{\rm f})\rightarrow\U(N_{\rm f}) \\ \text{AIII},\ \chi{\rm GUE}_\nu(n),\ \nu\in\mathbb{Z}\end{array}}$		&		$\overset{\ }{\begin{array}{c} [\gamma^{(5)},\mathcal{D}]_+=[\mathcal{C},\mathcal{D}]_+=[\mathcal{C},\gamma^{(5)}]_+=0,\\ \mathcal{C}^2= \eins \\ \USp(2N_{\rm f})\times\USp(2N_{\rm f})\rightarrow\USp(2N_{\rm f}) \\ \text{CI},\ {\rm GBOE}(n)\end{array}}$		\\ \hline 
  $8m+3$		
  &	$\overset{\ }{\begin{array}{c} [\mathcal{C},\mathcal{D}]_+=0,\ \mathcal{C}^2= -\eins \\ \Ort(2N_{\rm f})\rightarrow\Ort(N_{\rm f})\times\Ort(N_{\rm f})\\ \text{AII},\ {\rm GSE}(n)\end{array}}$	&	$\overset{\ }{\begin{array}{c} {\rm no\ further\ symmetries} \\ \U(2N_{\rm f})\rightarrow\U(N_{\rm f})\times\U(N_{\rm f})\\ \text{A},\ {\rm GUE}(n)\end{array}}$	&	$\overset{\ }{\begin{array}{c} [\mathcal{C},\mathcal{D}]_+=0,\ \mathcal{C}^2=\eins \\ \USp(4N_{\rm f})\rightarrow\USp(2N_{\rm f})\times\USp(2N_{\rm f})\\ \text{AI},\ {\rm GOE}(n)\end{array}}$	\\ \hline
  $8m+4$		
  &	$\overset{\ }{\begin{array}{c} [\gamma^{(5)},\mathcal{D}]_+=[\mathcal{C},\mathcal{D}]_-=[\mathcal{C},\gamma^{(5)}]_-=0,\\ \mathcal{C}^2= -\eins \\ \U(2N_{\rm f})\rightarrow\Ort(2N_{\rm f}) \\ \text{CII},\ \chi{\rm GSE}_\nu(n),\ \nu\in\mathbb{Z} \end{array}}$	 &	$\overset{\ }{\begin{array}{c} {[\gamma^{(5)},\mathcal{D}]_+=0} \\ \U(N_{\rm f})\times\U(N_{\rm f})\rightarrow\U(N_{\rm f}) \\ \text{AIII},\ \chi{\rm GUE}_\nu(n),\ \nu\in\mathbb{Z}\end{array}}$	&	$\overset{\ }{\begin{array}{c} [\gamma^{(5)},\mathcal{D}]_+=[\mathcal{C},\mathcal{D}]_-=[\mathcal{C},\gamma^{(5)}]_-=0,\\ \mathcal{C}^2= \eins \\ \U(2N_{\rm f})\rightarrow\USp(2N_{\rm f}) \\ \text{B$\mid$DI},\ \chi{\rm GOE}_\nu(n),\ \nu\in\mathbb{Z}\end{array}}$ 	\\ \hline
  $8m+5$		
  &	$\overset{\ }{\begin{array}{c} [\mathcal{C},\mathcal{D}]_-=0,\ \mathcal{C}^2= -\eins \\ \USp(2N_{\rm f})\rightarrow\U(N_{\rm f}) \\ \text{C},\ {\rm GASE}(n) \end{array}}$	&	$\overset{\ }{\begin{array}{c} {\rm no\ further\ symmetries} \\ \U(2N_{\rm f})\rightarrow\U(N_{\rm f})\times\U(N_{\rm f})\\ \text{A},\ {\rm GUE}(n)\end{array}}$		&		$\overset{\ }{\begin{array}{c} [\mathcal{C},\mathcal{D}]_-=0,\ \mathcal{C}^2= \eins \\ \Ort(2N_{\rm f})\rightarrow\U(N_{\rm f}) \\ \text{B$\mid$D},\ {\rm GAOE}_\nu(n),\ \nu\in\mathbb{Z}_2\end{array}}$	\\ \hline
  $8m+6$		
  &		$\overset{\ }{\begin{array}{c} [\gamma^{(5)},\mathcal{D}]_+=[\mathcal{C},\mathcal{D}]_+=[\mathcal{C},\gamma^{(5)}]_+=0,\\ \mathcal{C}^2= \eins \\ \USp(2N_{\rm f})\times\USp(2N_{\rm f})\rightarrow\USp(2N_{\rm f}) \\ \text{CI},\ {\rm GBOE}(n)\end{array}}$		&	$\overset{\ }{\begin{array}{c} {[\gamma^{(5)},\mathcal{D}]_+=0} \\ \U(N_{\rm f})\times\U(N_{\rm f})\rightarrow\U(N_{\rm f}) \\ \text{AIII},\ \chi{\rm GUE}_\nu(n),\ \nu\in\mathbb{Z}\end{array}}$		&	$\overset{\ }{\begin{array}{c} [\gamma^{(5)},\mathcal{D}]_+=[\mathcal{C},\mathcal{D}]_+=[\mathcal{C},\gamma^{(5)}]_+=0,\\ \mathcal{C}^2= -\eins \\ \Ort(2N_{\rm f})\times\Ort(2N_{\rm f})\rightarrow\Ort(2N_{\rm f}) \\ \text{DIII},\ {\rm GBSE}_\nu(n),\ \nu\in\mathbb{Z}_2\end{array}}$		\\ \hline
  $8m+7$		
  &	$\overset{\ }{\begin{array}{c} [\mathcal{C},\mathcal{D}]_+=0,\ \mathcal{C}^2=\eins \\ \USp(4N_{\rm f})\rightarrow\USp(2N_{\rm f})\times\USp(2N_{\rm f})\\ \text{AI},\ {\rm GOE}(n)\end{array}}$		&	$\overset{\ }{\begin{array}{c} {\rm no\ further\ symmetries} \\ \U(2N_{\rm f})\rightarrow\U(N_{\rm f})\times\U(N_{\rm f})\\ \text{A},\ {\rm GUE}(n)\end{array}}$		&	$\overset{\ }{\begin{array}{c} [\mathcal{C},\mathcal{D}]_+=0,\ \mathcal{C}^2= -\eins \\ \Ort(2N_{\rm f})\rightarrow\Ort(N_{\rm f})\times\Ort(N_{\rm f})\\ \text{AII},\ {\rm GSE}(n)\end{array}}$		\\ \hline
\end{tabular}
\caption{
Classification of QCD-like continuum  theories with respect to their spontaneous breaking of their flavor symmetry group like chiral symmetries, see~\cite{contclass}. In the first line of each entry the symmetries of the continuum Dirac operator $\mathcal{D}=-\mathcal{D}^\dagger$ regarding the antiunitary operator $\mathcal{C}$ and the matrix $\gamma^{(5)}$, if applicable, are shown. In the second line we show the symmetry breaking pattern but without taking into account the anomalous symmetry breaking of the axial symmetry group. The space-time dimension $d$ has to be larger than two for these patterns while $d=2$ does not necessarily exhibit a spontaneous symmetry breaking because of the Mermin-Wagner-Coleman theorem. However it was observed in numerical simulations that some QCD-like theories might show a symmetry breaking in two dimensions, see \cite{2DstagQCD}. In the third line of the entries we recall the symmetry class of the Cartan scheme~\cite{threefoldway,tenfoldway} and the corresponding random matrix model exhibiting the same spectral statistics for the lowest eigenvalues as the corresponding Dirac operator, see table~\ref{tab:RMT}. The indices $n$ and $\nu$ determine the dimension of the random matrix and play the roles of the volume and the topological charge.
}\label{tab:cont-class}
\end{table}

\begin{table} 
\begin{tabular}{|c||c|c|c|}
  \hline
  \diagbox{\ $d_{\rm red}=d-N_{\rm L}$\hspace*{-0.4cm}}{\hspace*{-0.6cm} Theory }
                 & real representation & complex representation & quaternion representation   \\
  \hline\hline
  $8m$			
  &	$\overset{\ }{\begin{array}{c} \U(2d_{\rm tri}N_{\rm f})\\ \downarrow\\ \USp(2d_{\rm tri}N_{\rm f}) \\ \text{BI},\ \chi{\rm GOE}_0(n)\end{array}}$		&	$\overset{\ }{\begin{array}{c} \U(d_{\rm tri}N_{\rm f})\times\U(d_{\rm tri}N_{\rm f}) \\ \downarrow\\ \U(d_{\rm tri}N_{\rm f}) \\ \text{AIII},\ \chi{\rm GUE}_0(n) \end{array}}$		&	$\overset{\ }{\begin{array}{c} \U(2d_{\rm tri}N_{\rm f}) \\ \downarrow \\ \Ort(2d_{\rm tri}N_{\rm f}) \\ \text{CII},\ \chi{\rm GSE}_0(n) \end{array}}$			\\ \hline
  $8m+1$		
  &	$\overset{\ }{\begin{array}{c} \Ort(2d_{\rm tri}N_{\rm f}) \\ \downarrow \\ \U(d_{\rm tri}N_{\rm f}) \\ \text{B},\ {\rm GAOE}_0(n) \end{array}}$	 &	$\overset{\ }{\begin{array}{c} \U(2d_{\rm tri}N_{\rm f}) \\ \downarrow \\ \U(d_{\rm tri}N_{\rm f})\times\U(d_{\rm tri}N_{\rm f})\\ \text{A},\ {\rm GUE}(n) \end{array}}$		&	$\overset{\ }{\begin{array}{c} \USp(2d_{\rm tri}N_{\rm f})\\ \downarrow \\ \U(d_{\rm tri}N_{\rm f}) \\ \text{C},\ {\rm GASE}(n) \end{array}}$	\\ \hline  
  $8m+2$		
  &	$\overset{\ }{\begin{array}{c} \Ort(2d_{\rm tri}N_{\rm f})\times\Ort(2d_{\rm tri}N_{\rm f}) \\ \downarrow \\ \Ort(2d_{\rm tri}N_{\rm f}) \\ \text{DIII},\ {\rm GBSE}_0(n) \end{array}}$	&	$\overset{\ }{\begin{array}{c} \U(d_{\rm tri}N_{\rm f})\times\U(d_{\rm tri}N_{\rm f}) \\ \downarrow \\ \U(d_{\rm tri}N_{\rm f}) \\ \text{AIII},\ \chi{\rm GUE}_0(n) \end{array}}$		&		$\overset{\ }{\begin{array}{c} \USp(2d_{\rm tri}N_{\rm f})\times\USp(2d_{\rm tri}N_{\rm f}) \\ \downarrow \\ \USp(2d_{\rm tri}N_{\rm f}) \\ \text{CI},\ {\rm GBOE}(n)\end{array}}$		\\ \hline 
  $8m+3$		
  &	$\overset{\ }{\begin{array}{c} \Ort(2d_{\rm tri}N_{\rm f}) \\ \downarrow \\ \Ort(d_{\rm tri}N_{\rm f})\times\Ort(d_{\rm tri}N_{\rm f})\\ \text{AII},\ {\rm GSE}(n)\end{array}}$	&	$\overset{\ }{\begin{array}{c} \U(2d_{\rm tri}N_{\rm f}) \\ \downarrow \\ \U(d_{\rm tri}N_{\rm f})\times\U(d_{\rm tri}N_{\rm f})\\ \text{A},\ {\rm GUE}(n)\end{array}}$	&	$\overset{\ }{\begin{array}{c} \USp(4d_{\rm tri}N_{\rm f}) \\ \downarrow \\ \USp(2d_{\rm tri}N_{\rm f})\times\USp(2d_{\rm tri}N_{\rm f})\\ \text{AI},\ {\rm GOE}(n)\end{array}}$	\\ \hline
  $8m+4$		
  &	$\overset{\ }{\begin{array}{c} \U(2d_{\rm tri}N_{\rm f}) \\ \downarrow \\ \Ort(2d_{\rm tri}N_{\rm f}) \\ \text{CII},\ \chi{\rm GSE}_0(n) \end{array}}$	 &	$\overset{\ }{\begin{array}{c} \U(d_{\rm tri}N_{\rm f})\times\U(d_{\rm tri}N_{\rm f}) \\ \downarrow \\ \U(d_{\rm tri}N_{\rm f}) \\ \text{AIII},\ \chi{\rm GUE}_0(n) \end{array}}$	&	$\overset{\ }{\begin{array}{c} \U(2d_{\rm tri}N_{\rm f}) \\ \downarrow \\ \USp(2d_{\rm tri}N_{\rm f}) \\ \text{BI},\ \chi{\rm GOE}_0(n) \end{array}}$ 	\\ \hline
  $8m+5$		
  &	$\overset{\ }{\begin{array}{c} \USp(2d_{\rm tri}N_{\rm f}) \\ \downarrow \\ \U(d_{\rm tri}N_{\rm f}) \\ \text{C},\ {\rm GASE}(n) \end{array}}$	&	$\overset{\ }{\begin{array}{c} \U(2d_{\rm tri}N_{\rm f}) \\ \downarrow \\ \U(d_{\rm tri}N_{\rm f})\times\U(d_{\rm tri}N_{\rm f})\\ \text{A},\ {\rm GUE}(n)\end{array}}$		&		$\overset{\ }{\begin{array}{c} \Ort(2d_{\rm tri}N_{\rm f}) \\ \downarrow \\ \U(d_{\rm tri}N_{\rm f}) \\ \text{B},\ {\rm GAOE}_0(n) \end{array}}$	\\ \hline
  $8m+6$		
  &		$\overset{\ }{\begin{array}{c} \USp(2d_{\rm tri}N_{\rm f})\times\USp(2d_{\rm tri}N_{\rm f}) \\ \downarrow \\ \USp(2d_{\rm tri}N_{\rm f}) \\ \text{CI},\ {\rm GBOE}(n)\end{array}}$		&	$\overset{\ }{\begin{array}{c} \U(d_{\rm tri}N_{\rm f})\times\U(d_{\rm tri}N_{\rm f}) \\ \downarrow \\ \U(d_{\rm tri}N_{\rm f}) \\ \text{AIII},\ \chi{\rm GUE}_0(n) \end{array}}$		&	$\overset{\ }{\begin{array}{c} \Ort(2d_{\rm tri}N_{\rm f})\times\Ort(2d_{\rm tri}N_{\rm f}) \\ \downarrow \\ \Ort(2d_{\rm tri}N_{\rm f}) \\ \text{DIII},\ {\rm GBSE}_0(n) \end{array}}$		\\ \hline
  $8m+7$		
  &	$\overset{\ }{\begin{array}{c} \USp(4d_{\rm tri}N_{\rm f}) \\ \downarrow \\ \USp(2d_{\rm tri}N_{\rm f})\times\USp(2d_{\rm tri}N_{\rm f})\\ \text{AI},\ {\rm GOE}(n) \end{array}}$		&	$\overset{\ }{\begin{array}{c} \U(2d_{\rm tri}N_{\rm f}) \\ \downarrow \\ \U(d_{\rm tri}N_{\rm f})\times\U(d_{\rm tri}N_{\rm f})\\ \text{A},\ {\rm GUE}(n) \end{array}}$		&	$\overset{\ }{\begin{array}{c} \Ort(2d_{\rm tri}N_{\rm f}) \\ \downarrow \\ \Ort(d_{\rm tri}N_{\rm f})\times\Ort(d_{\rm tri}N_{\rm f})\\ \text{AII},\ {\rm GSE}(n) \end{array}}$		\\ \hline
\end{tabular}
\caption{
The symmetry breaking patterns of the naive lattice Dirac operator for the three kinds of theories with a dimension $d>2$ and the corresponding Cartan class~\cite{threefoldway,tenfoldway} and the random matrix theory, see table~\ref{tab:RMT} and appendix~\ref{sec:RMT}, with which the Monte Carlo simulations have to be compared. Instead of the dimension $d$ of the continuum theory one obtains the symmetry of a shifted dimension $d_{\rm red}=d-N_{\rm L}$. The staggered Dirac operator corresponds to $d_{\rm red}=0=8m$ though the degeneracy $d_{\rm tri}=2^{\lfloor (N_{\rm L}+[d+1]_2)/2\rfloor}$ will not be present for this operator. The degeneracy is also the reason why the number of flavors is increased to $d_{\rm tri}N_{\rm f}$ compared to the continuum theory, cf. table~\ref{tab:cont-class}. Note that the topological charge $\nu$ vanishes because no zero modes are present in the naive Dirac operator with $d>2$.
}\label{tab:lat-class}
\end{table}

\begin{table}[t!]
\centering
\rotatebox{90}{
\begin{tabular}{|c|c|c|c|c|c|c|c|c|}
  \hline
  RMT
                 & $\begin{array}{c} \text{Abbreviation for} \\ \text{Gaussian ensemble}\end{array}$ & $\begin{array}{c} {\rm Cartan} \\ {\rm class}\end{array}$ & Random matrix $H$ & $\quad\beta_{\rm D}\quad$ & $\alpha_{\rm D}$ & $\nu_{\rm D}$ & $\begin{array}{c} \text{chiral} \\ \text{pair} \end{array}$  & $\begin{array}{c} \text{generic} \\ \text{zeros} \end{array}$ \\
  \hline\hline
  $\begin{array}{c} \text{Hermitian matrices} \\ \text{Lie algebra of }\U(n) \end{array}$		&	GUE$(n)$	&	A	&	$H=H^\dagger\in\mathbb{C}^{n\times n},\ n\in\mathbb{N}$ & $2$  & $0$ & $0$ & No & $0$ \\ \hline
  real symmetric matrices		&	GOE$(n)$	&	AI	&	$\overset{\ }{H=H^T=H^*\in\mathbb{R}^{n\times n}},\ n\in\mathbb{N}$ & $1$  & $0$ & $0$	& No & $0$ \\ \hline
   $\begin{array}{c} \text{Hermitian} \\ \text{self-dual matrices} \end{array}$		&	GSE$(n)$	&	AII	&	$\overset{\ }{\begin{array}{c} \displaystyle H=\theta_2H^T\theta_2=\theta_2H^*\theta_2\\
   H\in\mathbb{C}^{2n\times 2n},\ n\in\mathbb{N}\end{array}}$ & $4$  & $0$ & $0$	& No & $0$ \\ \hline
  $\begin{array}{c} \text{imaginary antisymmetric}\\ \text{matrices} \\ \text{Lie algebra of }\Ort(n) \end{array}$	&	GAOE$_\nu(n)$	&	B$\mid$D	&	$\overset{\ }{\begin{array}{c} \displaystyle H=-H^T=-H^*\\ \displaystyle H\in\imath\mathbb{R}^{(2n+\nu)\times(2n+\nu)},\\ \displaystyle\ n\in\mathbb{N}, \nu=0,1\end{array}}$ & $2$  & $\quad 2\nu=0,2\quad$ & $\quad\displaystyle\frac{\nu-1}{2}=\pm\frac{1}{2}\quad$	 & Yes & $\nu=0,1$ \\ \hline
  $\begin{array}{c} \text{Hermitian anti-self-dual}\\ \text{matrices} \\ \text{Lie algebra of }\USp(2n) \end{array}$	&	GASE$(n)$	&		C &	$\overset{\ }{\begin{array}{c} \displaystyle H=-\theta_2H^T\theta_2=-\theta_2H^*\theta_2\\ H\in\mathbb{C}^{2n\times2n},\ n\in\mathbb{N}\end{array}}$ & $2$  & $2$ & $\displaystyle\frac{1}{2}$ & Yes & $0$ \\ \hline
  chiral Hermitian matrices		&	$\chi$GUE$_\nu(n)$	&	AIII	&	$\overset{\ }{\begin{array}{c} \displaystyle H=\left[\begin{array}{cc} 0 & W \\ W^\dagger & 0 \end{array}\right],\\ \displaystyle W\in\mathbb{C}^{n\times(n+\nu)},\ n,\nu\in\mathbb{N}\end{array}}$ & $2$  & $2\nu+1$ & $\nu$ & Yes & $\nu$ \\ \hline
  $\begin{array}{c} \text{chiral real}\\ \text{symmetric matrices}\end{array}$ 	&	$\chi$GOE$_\nu(n)$	&	B$\mid$DI	&	$\overset{\ }{\begin{array}{c} \displaystyle H=\left[\begin{array}{cc} 0 & W \\ W^\dagger & 0 \end{array}\right],\\ \displaystyle W=W^*\in\mathbb{R}^{n\times(n+\nu)},\ n,\nu\in\mathbb{N}\end{array}}$ & $1$  & $\nu$ & $\nu$ & Yes & $\nu$ \\ \hline
   $\begin{array}{c} \text{chiral Hermitian} \\ \text{self-dual matrices} \end{array}$		&	$\chi$GSE$_\nu(n)$	&	CII	&	$\overset{\ }{\begin{array}{c} \displaystyle H=\left[\begin{array}{cc} 0 & W \\ W^\dagger & 0 \end{array}\right],\\ W=\theta_2W^*\theta_2\in\mathbb{C}^{2n\times2(n+\nu)},\\ n,\nu\in\mathbb{N}\end{array}}$ & $4$ & $4\nu+3$ & $\nu$ & Yes & $2\nu$ \\ \hline
  $\begin{array}{c} \text{symmetric Bogolyubov-} \\ \text{de Gennes matrices} \end{array}$	&	GBOE$(n)$	&	CI	&	$\overset{\ }{\begin{array}{c} \displaystyle H=\left[\begin{array}{cc} 0 & W \\ W^\dagger & 0 \end{array}\right],\\ \displaystyle W=W^T\in\mathbb{C}^{n\times n},\ n\in\mathbb{N} \end{array}}$ & $1$ & $1$  & $1$ & Yes & $0$ \\ \hline
   	$\begin{array}{c} \text{antisymmetric Bogolyubov-} \\ \text{de Gennes matrices} \end{array}$	&	GBSE$_\nu(n)$	&	DIII	&	$\overset{\ }{\begin{array}{c} \displaystyle H=\left[\begin{array}{cc} 0 & W \\ W^\dagger & 0 \end{array}\right],\\ \displaystyle W=-W^T\in\mathbb{C}^{(2n+\nu)\times (2n+\nu)},\\ \displaystyle n\in\mathbb{N},\ \nu=0,1 \end{array}}$	 & $4$  & $4\nu+1$ & $\displaystyle\frac{\nu-1}{2}=\pm\frac{1}{2}$ & Yes & $2\nu=0,2$ \\ \hline
\end{tabular}}
\caption{
The ten Gaussian random matrix models corresponding to the ten symmetries of the Cartan classification scheme~\cite{threefoldway,tenfoldway}. We want to underline that not all of the abbreviations for these ensemble are standard. The last five classes are the chiral models while the first five exhibit no chirality but may have ``chiral pairs" of eigenvalues $(\lambda,-\lambda)$, see seventh columns. Shown are their explicit matrix representations (fourth column), the Dyson index $\beta_{\rm D}$ (fifth column) and the exponent of the level repulsion from the origin $\alpha_{\rm D}$ (sixth columns as well as the number of generic zero modes. Please note that some ensembles only differ in their spectral properties via subtleties like the number of the zero modes. The indices $\beta_{\rm D}$, $\alpha_{\rm D}$ and $\nu_{\rm D}$ are needed for the analytical random matrix results~\eqref{jpdf-a}, \eqref{jpdf-b}, \eqref{level-dens} and \eqref{Wigner}. 
}\label{tab:RMT}
\end{table}

\begin{figure}[t!]
\centerline{\includegraphics[width=1\textwidth]{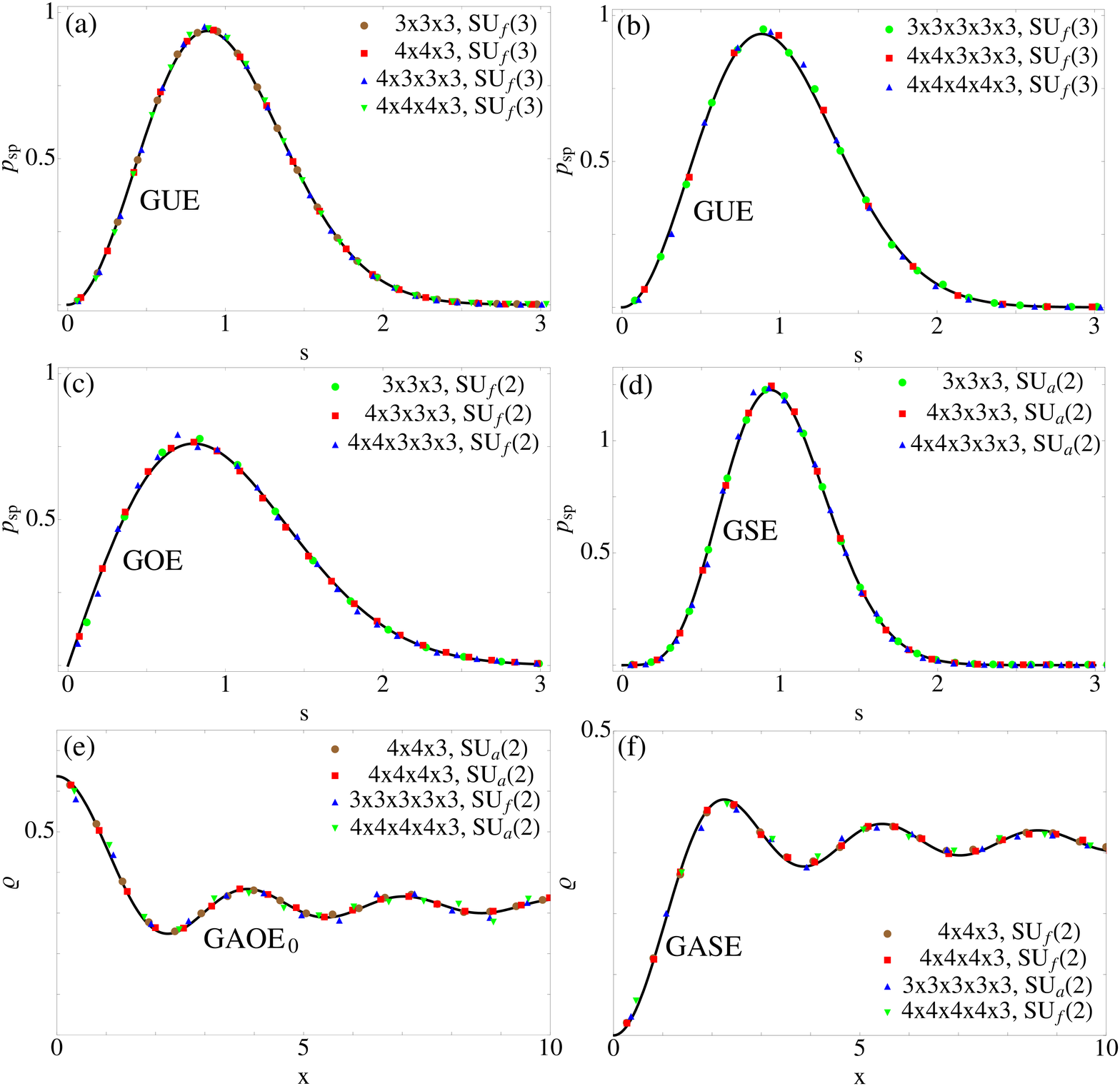}}
\caption{Comparison of the random matrix theory predictions (black smooth curves) with lattice simulations of three-, four- and five-dimensional quenched naive Dirac operators (coloured symbols) in the strong coupling limit. The abbreviations GUE, GOE, $\ldots$ refer to the corresponding random matrix ensemble listed in table~\ref{tab:RMT} and the abbreviations $\SU_{f}(N_{\rm c})$ and $\SU_{a}(N_{\rm c})$ stand for the fundamental and adjoint representation of the group $\SU(N_{\rm c})$. In the first plots we show the level spacing distribution $p_{\rm sp}(s)$ which is given by Wigner's surmise~\eqref{Wigner}. The lattice data for these plots were normalized to unity for the zeroth and first moment. For these plots we took into account $19$ eigenvalues ($=18$ level spacings) per configuration of the Dirac operator about the origin. The last two plots show the microscopic level density $\varrho(x)$, see eq.~\eqref{level-dens}. For those plots the lattice data was rescaled via a $\chi^2$-fitting. We simulated $10^5$ configurations for the three- and four-dimensional lattices. For the five-dimensional lattices we generated below $10^4$ configurations such that the statistical variance will be below five percent.
}
\label{fig1}
\end{figure}

\begin{figure}[t!]
\centerline{\includegraphics[width=1\textwidth]{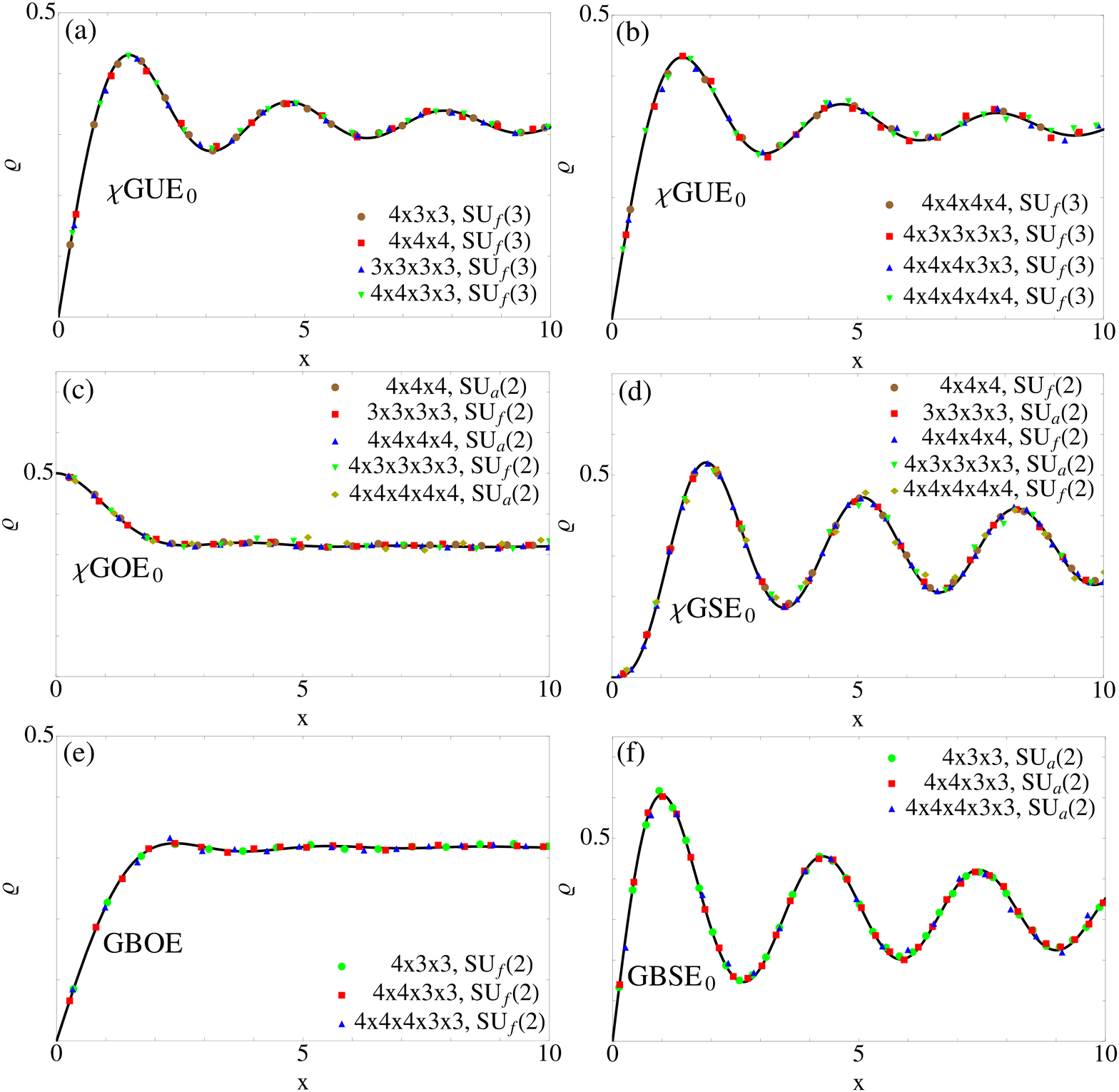}}
\caption{Continuation of the list of comparisons of figure~\ref{fig1} between random matrix theory predictions (black smooth curves) and lattice simulations of three-, four- and five-dimensional quenched naive Dirac operators (coloured symbols) in the strong coupling limit. In all six plots we consider the microscopic level density $\varrho(x)$, see eq.~\eqref{level-dens}, only. As in figure~\ref{fig1} we applied a $\chi^2$-fitting on the lattice data which consist of $10^5$ configurations for the three- and four-dimensional lattices and the number of configurations varies ($<15000$) for the five-dimensional lattices.
}
\label{fig2}
\end{figure}
\end{widetext}


\newpage
\begin{thebibliography}{10}

\bibitem{technicolor}
 E. Farhi and L. Susskind: \textit{Technicolor}, Phys. Rept. {\bf 74}, 277 (1981).
 
 \bibitem{SUSYQCD}
M. Shin-Mura and K. Yamawaki:  \textit{Chiral Symmetry Breakings in Supersymmetric QCD}, Progress of Theoretical Physics {\bf 71}, 1036 (1984). 

\bibitem{staggered}
 L. Susskind: \textit{Lattice fermions}, Phys. Rev. D {\bf 16}, 3031 (1977).
 
 \bibitem{Wilson}
K. Wilson: \textit{New Phenomena In Subnuclear Physics}, edited by A. Zichichi, Plenum Press: New York, {\bf 69}
(1977).
 
 \bibitem{Twisted}
R. Frezzotti, P. A. Grassi, S. Sint, and P. Weisz: \textit{Lattice QCD with a chirally twisted mass term},  Journal of High Energy Physics  {\bf JHEP08}, 058 (2001) [arXiv:hep-lat/0101001].
 
 \bibitem{Overlap}
R. Narayanan, H. Neuberger: \textit{A construction of lattice chiral gauge theories}, Nucl. Phys. B {\bf 443}, 305 (1995). 
 
 \bibitem{Domain}
 D. B. Kaplan: \textit{A method for simulating chiral fermions on the lattice}, Phys. Lett. B {\bf 288}, 342 (1992).
 
\bibitem{4DstagQCD.a}
P.H. Damgaard, U.M. Heller, R. Niclasen, and B. Svetitsky: \textit{Patterns of spontaneous chiral symmetry breaking in vector like gauge theories}, Nucl. Phys. B {\bf 633}, 97 (2002) [arXiv:hep-lat/0110028]. 
 
 \bibitem{4DstagQCD.b}
 F. Bruckmann, S. Keppeler, M. Panero, and T. Wettig: \textit{Polyakov loops and SU(2) staggered Dirac spectra}, PoS {\bf LATTICE 2007}, 274 (2007) [arXiv:0802.0662 [hep-lat]]; \textit{Polyakov loops and spectral properties of the staggered Dirac operator}, Phys. Rev. D {\bf 78}, 034503 (2008) [arXiv:0804.3929 [hep-lat]].
 
 \bibitem{3DstagQCD.a}
 P. H. Damgaard, U. M. Heller, A. Krasnitz, and T. Madsen: \textit{A Quark-Antiquark Condensate in Three-Dimensional QCD}, Phys. Lett. B {\bf 440}, 129  (1998) [arXiv:hep-lat/9803012].
 
 \bibitem{3DstagQCD.b}
 P. Bialas, Z. Burda, and B. Petersson: \textit{Random matrix model for QCD$_3$ staggered fermions}, Phys. Rev. D {\bf 83}, 014507 (2011) [arXiv:1006.0360 [hep-lat]].
 
 \bibitem{2DstagQCD}
 M. Kieburg, J. J. M. Verbaarschot, and S. Zafeiropoulos: \textit{A classification of 2-dim Lattice Theory}, PoS {\bf LATTICE 2013}, 337 (2013) [arXiv:1310.6948 [hep-lat]];  \textit{Dirac Spectra of 2-dimensional QCD-like theories}, Phys. Rev. D {\bf 90}, 085013 (2014) [arXiv:1405.0433 [hep-lat]].
 
\bibitem{BanksCasher}
T. Banks and A. Casher: \textit{Chiral symmetry breaking in confining theories}, Nucl. Phys. B {\bf 169}, 103 (1980). 

\bibitem{GasserLeutwyler}
J. Gasser and H. Leutwyler: \textit{Thermodynamics of chiral symmetry}, Phys. Lett. B {\bf 188}, 477 (1987).

\bibitem{LeutwylerSmilga}
H. Leutwyler and A. Smilga: \textit{Spectrum of Dirac operator and role of winding number in QCD}, Phys. Rev. D {\bf 46}, 5607 (1992).

\bibitem{ShuVer93}
  E.~V.~Shuryak, J.~J.~M.~Verbaarschot: \textit{Random matrix theory and spectral sum rules for the Dirac operator in QCD},
  Nucl.\ Phys.\  A {\bf 560}, 306-320 (1993) [arXiv:hep-th/9212088].

\bibitem{Peskin}
M. Peskin:  \textit{The alignment of the vacuum in theories of technicolor},
Nuc. Phys. B {\bf 175}, 197 (1980); \textit{Chiral Symmetry and Chiral Symmetry Breaking},
in Les Houches 1982: Recent advances in field theory and statistical
mechanics, Amsterdam, North-Holland (1984).

\bibitem{Preskill}
J. Preskill: \textit{Subgroup alignment in hypercolor theories}, Nuc.
Phys. B {\bf 177}, 21 (1981).
  
\bibitem{Verthreefold}
J. J. M. Verbaarschot: \textit{The spectrum of the QCD Dirac operator and chiral random matrix theory: the threefold way}, Phys. Rev. Lett. {\bf 72}, 2531 (1994) [arXiv:hep-th/9401059].

\bibitem{contclass}
R. DeJonghe, K. Frey, and T. Imbo: \textit{Bott Periodicity and Realizations of Chiral Symmetry in Arbitrary Dimensions}, Phys. Lett. B {\bf 718}, 603 (2012) [arXiv:1207.6547 [hep-th]].

\bibitem{Bott}
 R. Bott: \textit{The stable homotopy of the classical groups}, Ann. Math. {\bf 70}, 313 (1959).
 
\bibitem{Clifford}
D. J. H. Garling: \textit{Clifford Algebras: An Introduction}, London Mathematical Society Texts {\bf 78}, Cambridge University Press, Cambridge (2011).

\bibitem{threefoldway}
F. J. Dyson: \textit{The Threefold Way. Algebraic Structure of Symmetry Groups and Ensembles in Quantum Mechanics}, J. Math. Phys. {\bf 3}, 1199 (1962).

\bibitem{tenfoldway}
M. R. Zirnbauer: \textit{Riemannian symmetric superspaces and their origin in random matrix theory}, J. Math. Phys. {\bf 37}, 4986 (1996) [arXiv:math-ph/9808012]; A. Altland, M.R. Zirnbauer: \textit{Novel Symmetry Classes in Mesoscopic Normal-Superconducting Hybrid Structures}, Phys. Rev. B {\bf 55}, 1142 (1997) [arXiv:cond-mat/9602137].

\bibitem{topInsu}
A. P. Schnyder, S. Ryu, A. Furusaki, and A. W. W. Ludwig: \textit{Classification of topological insulators and superconductors in three spatial dimensions}, Phys. Rev. B {\bf 78} , 195125 (2008) [arXiv:0803.2786 [cond-mat.mes-hall]].

\bibitem{MerminWagner}
N. D. Mermin and H. Wagner: \textit{Absence of Ferromagnetism or Antiferromagnetism in One- or Two-Dimensional Isotropic Heisenberg Models}, Phys. Rev. Lett. {\bf 17}, 1133 (1966); Erratum Phys. Rev. Lett. {\bf 17}, 1307 (1966).

\bibitem{coleman}
S. Coleman: \textit{There are no Goldstone bosons in two dimensions}, Comm. Math. Phys. {\bf 31}, 259 (1973).

\bibitem{doubler}
S. Chandrasekharan and U.-J. Wiese: \textit{An Introduction to chiral symmetry on the lattice}, Prog. Part. Nucl. Phys. {\bf 53} 373 (2004) [arXiv:hep-lat/0405024].

\bibitem{Schur}
I. Schur: \textit{Neue Begr\"undung der Theorie der Gruppencharaktere}, Sitzungsberichte der K\"oniglich Preu\ss ischen Akademie der Wissenschaften zu Berlin, 406 (1905).

\bibitem{Majorana}
M. Leijnse and K. Flensberg: \textit{Introduction to topological superconductivity and Majorana fermions}, Semicond. Sci. Technol. {\bf 27}, 124003 (2012) [arXiv:1206.1736 [cond-mat.mes-hall]].

\bibitem{Follana}
E. Follana, A. Hart, and C. T. H. Davies: \textit{The Index Theorem and Universality Properties of the Low-lying Eigenvalues of Improved Staggered Quarks}, Phys. Rev. Lett. {\bf 93}, 241601 (2004) [arXiv:hep-lat/0406010].


\bibitem{Leveldensity}
D. Ivanov: \textit{The supersymmetric technique for random-matrix ensembles with zero eigenvalues}, J. Math. Phys. {\bf 43}, 126 (2002) [arXiv:cond-mat/0103137].

\bibitem{Wigner}
E. P. Wigner: \textit{Results and Theory of Resonance Absorbtion}, Conference Proceeding of Conference on Neutron Physics by Time-of-Flight held at Oak Ridge National Laboratory in Gatlinburg (Tennessee/USA) 1956, 59 (1957).

\bibitem{Osborn}
J. Osborn: \textit{Taste breaking in staggered fermions from random matrix theory}, Nucl. Phys. Suppl. {\bf 129}, 886 (2004) [arXiv:hep-lat/0309123]; \textit{Staggered chiral random matrix theory}, Phys. Rev. D {\bf 83}, 034505 (2011) [arXiv:1012.4837 [hep-lat]]; \textit{Chiral random matrix theory for staggered fermions}, PoS {\bf Lattice 2011}, 110 (2011) [arXiv:1204.5497 [hep-lat]].

\bibitem{Dietz}
B. Dietz and F. Haake: \textit{Taylor and Padé analysis of the level spacing distributions of
random-matrix ensembles}, Z. Phys. B-Condensed Matter {\bf 80}, 153 (1990).

\end{thebibliography}
\end{document}